\documentclass[12pt,a4paper]{article}
\usepackage[dvips]{psfrag}
\usepackage{cite}

\usepackage{braket}
\usepackage[normalem]{ulem}

\usepackage{textcomp}
\usepackage{amsmath,amsthm,amssymb,mathrsfs,cases}
\usepackage{amsfonts,graphicx,lscape}

\def\sech{\mbox{\,sech}}

\newcommand{\vt}[1]{\mbox{\boldmath$#1$}}
\newcommand{\sca}[2]{\langle #1, #2 \rangle}

\numberwithin{equation}{section} 
\newtheorem{theorem}{Theorem}[section]
\newtheorem{proposition}[theorem]{Proposition}

\def\beqa{\begin{eqnarray}}
\def\enqa{\end{eqnarray}}
\def\beq{\begin{equation}}
\def\enq{\end{equation}}

\begin{document}
\title{
\vspace{-9mm}
%
Multisoliton 
solutions  
of the 
vector nonlinear Schr\"odinger 
equation 
(Kulish--Sklyanin model)
and 
the vector mKdV equation 
}
\author{Takayuki \textsc{Tsuchida}
}
\maketitle
\begin{abstract} 
%
There exist two 
natural 
vector 
generalizations
of the 
completely 
integrable 
nonlinear Schr\"odinger (NLS) equation
in \mbox{$1+1$} dimensions:\ 
the well-known 
Manakov model and 
the lesser-known 
Kulish--Sklyanin model. 
In this paper, we propose 
a binary 
Darboux 
(or 
Zakharov--Shabat
dressing) 
transformation 
that can be 
directly 
applied 
to 
the 
Kulish--Sklyanin model. 
By deriving 
a simple 
closed 
expression 
for 
iterations of the binary Darboux transformation, 
we obtain 
an explicit 
formula 
for the 
$N$-soliton solution 
of the 
Kulish--Sklyanin model 
under 
vanishing 
boundary conditions. 
Because the third-order symmetry of the 
vector NLS equation 
can 
be reduced to 
a vector generalization of 
the 
modified KdV (mKdV)
equation, 
we can also obtain 
multisoliton (or multi-breather) solutions 
of the vector mKdV equation 
in closed form. 
\end{abstract}
%

\newpage
\section{Introduction}
The cubic nonlinear Schr\"odinger (NLS) equation~\cite{ZS1,ZS2}
\begin{equation}
\label{NLS}
 \mathrm{i} q_t + q_{xx} +2 \sigma 
	|q|^2 q = 0, \quad 
\sigma = \mbox{$+1$} \; \mathrm{or} \; \mbox{$-1$}, 
\end{equation}
is a representative 
integrable system 
in \mbox{$1+1$} 
dimensions. 
The 
case \mbox{$\sigma = +1$} and the 
case \mbox{$\sigma = -1$} 
correspond to the self-focusing and self-defocusing NLS equation, 
respectively.  
The NLS equation 
can be generalized to a 
single vector 
equation 
involving 
the standard 
scalar product \mbox{$\sca{\,\cdot\,}{\,\cdot\,}$}
in two distinct ways 
while 
preserving the integrability~\cite{SWo01}; 
that is, 
the Manakov model~\cite{Mana74} 
\begin{equation}
\label{rMana}
 \mathrm{i} \vt{q}_t + \vt{q}_{xx} + 2\sca{\vt{q}}{\vt{q}^\ast}
	\vt{q} = \vt{0}
\end{equation}
and the Kulish--Sklyanin model~\cite{KuSk81}
\begin{equation}
\label{rKS}
 \mathrm{i} \vt{q}_t + \vt{q}_{xx} + 4 \sca{\vt{q}}{\vt{q}^\ast} \vt{q}
	- 2 \sca{\vt{q}}{\vt{q}} \vt{q}^\ast = \vt{0}. 
\end{equation}
Here, \mbox{$\vt{q}$}
is 
a 
vector dependent variable 
and 
the asterisk denotes the complex conjugation. 
For brevity, we write down only the self-focusing case here, 
but it is straightforward to extend 
these models 
to 
the self-defocusing or 
a mixed focusing-defocusing case~\cite{YO2,Ab78,New79,MMP81,Mak82,ZakShul82}. 
Note that these models often appear in some disguised forms; 
any invertible 
linear transformation 
can be applied to the vector 
\mbox{$\vt{q}$}, 
which mixes 
its 
components. 
The Kulish--Sklyanin model (\ref{rKS}) 
can be reduced to the Manakov model (\ref{rMana})
by setting 
\mbox{$\sca{\vt{q}}{\vt{q}}=0$},  
up to a trivial rescaling; 
this can be 
realized by 
restricting 
the components of 
\mbox{$\vt{q}$} as, {\it e.g.}, 
\[
\vt{q} = \left( q_1, \pm \mathrm{i} q_1, q_3, \pm \mathrm{i} q_3, \ldots, 
 q_{2m-1}, \pm \mathrm{i} q_{2m-1} 
\right). 
\]
This 
simple 
observation 
demonstrates 
that 
the explicit 
formula for the 
$N$-soliton solution of 
the Kulish--Sklyanin model (\ref{rKS}) 
and the vector soliton interactions thereof 
are 
highly nontrivial 
and 
more complicated than those 
for the Manakov model (\ref{rMana}) 
reported in~\cite{Mana74,Dub88,Tsuchi04,Caud2014}.  

Clearly, the 
Manakov model (\ref{rMana}) 
is 
obtained from 
the 
(generally rectangular) 
matrix generalization of the scalar NLS equation, 
i.e., the matrix NLS equation~\cite{ZS74}: 
\begin{equation}
\label{redNLS}
 \mathrm{i} Q_t + Q_{xx} + 2Q Q^\dagger Q = O, 
\end{equation}
as a special case. 
Here, the dagger denotes the Hermitian conjugation 
and the symbol $O$ 
is used to stress 
that this is a 
matrix equation. 
In contrast, 
the Kulish--Sklyanin model 
(\ref{rKS}) 
is obtained from the 
matrix NLS equation (\ref{redNLS})
through 
the
nontrivial reduction 
\[
Q = q_1 I + \sum_{j=1}^{2m-1} q_{j+1} e_j. 
\]
Here, 
$I$ is the 
identity matrix; 
\mbox{$\{ e_1, e_2, 
\ldots, e_{2m-1} \}$} 
are 
skew-Hermitian (\mbox{$e_j^\dagger = - e_j$})
matrices 
that 
form 
generators of 
the Clifford algebra, i.e., they 
satisfy 
the anticommutation relations:
\begin{equation}
\{ e_j , e_k \}_+
:=e_j e_k + e_k e_j = -2 \delta_{jk} I, 
\label{a-commute0}
\end{equation}
where $\delta_{jk}$ is 
the Kronecker delta. 
We require that 
\mbox{$\{ I, e_1, e_2, 
\ldots, e_{2m-1} \}$} 
are 
linearly independent. 

Because the ancestor model, 
the matrix NLS equation (\ref{redNLS}), 
can be solved using the inverse scattering method 
and the $N$-soliton solution can be 
written down 
explicitly, 
it 
is
straightforward to 
obtain the $N$-soliton solution of the Kulish--Sklyanin model (\ref{rKS}) 
through the 
reduction. 
However, 
the obtained expression is 
``non-classical" 
in the sense that it involves 
the generators of the Clifford algebra 
\mbox{$\{ e_1, e_2, \ldots, e_{2m-1} \}$} 
explicitly 
in a rather complicated manner; 
it is a highly nontrivial task to 
translate 
such a 
``non-classical" 
expression
into 
a 
more 
user-friendly 
``classical" 
expression 
not involving \mbox{$\{ e_1, e_2, \ldots, e_{2m-1} \}$}, using 
the anticommutation relations (\ref{a-commute0}). 
Indeed, this 
can be achieved 
for 
the one-
and two-soliton solutions, 
but 
not for the 
general 
$N$-soliton solution 
in practice. 

The main objective of this paper is to derive 
a 
simple 
closed 
expression 
for 
the general $N$-soliton solution of the Kulish--Sklyanin model (\ref{rKS}) 
without recourse to the 
$N$-soliton solution of 
the matrix NLS equation (\ref{redNLS}). 
To this end, 
we consider 
a nonstandard Lax representation
for the Kulish--Sklyanin model (\ref{rKS})~\cite{Adler94}, 
which does not involve 
the generators of the Clifford algebra 
\mbox{$\{ e_1, e_2, \ldots, e_{2m-1} \}$}, 
and 
apply a binary Darboux
(or Zakharov--Shabat
dressing)
transformation~\cite{ZS79,Sall82,Chud2,ZaMi80,MatSall91}. 
A peculiar structure 
of the binary Darboux transformation 
allows us to express 
an arbitrary number of its iterations 
in 
simple explicit form. 
Thus, by applying the 
$N$-fold binary Darboux transformation 
to the 
trivial 
zero solution, 
we obtain the bright 
$N$-soliton solution of the Kulish--Sklyanin model (\ref{rKS}) 
in closed form. 
Actually, 
the 
binary Darboux transformation can be 
applied to 
all 
the isospectral flows 
that belong to the same integrable hierarchy 
as the Kulish--Sklyanin model (\ref{rKS}).  
Among the 
higher 
flows of 
this integrable 
hierarchy, 
the third-order flow 
is 
particularly 
interesting 
because it 
simplifies 
to 
a vector analog of 
the modified KdV (mKdV) equation~\cite{Svi93,SviSok94}: 
\begin{align}
& \vt{q}_{y} 
+ 
\vt{q}_{xxx} + 6 \sca{\vt{q}}{\vt{q}}\vt{q}_x 
= \vt{0}
\label{vmKdV1}
\end{align}
under 
the reduction \mbox{$\vt{q}=\vt{q}^\ast$}. 
Thus, with a minor 
tune-up 
of the multifold binary Darboux transformation, 
we can 
obtain 
multisoliton solutions, 
multi-breather solutions 
and their mixtures of 
the vector mKdV equation (\ref{vmKdV1}). 

This paper is organized as follows. 
In section~2, we 
summarize 
two 
different 
Lax 
representations 
for the Kulish--Sklyanin model (\ref{rKS}) 
and make some remarks on its soliton solutions. 
In section~3, we 
propose 
the 
binary Darboux transformation
and 
apply 
its 
$N$-fold version 
to the Kulish--Sklyanin model (\ref{rKS}) 
to obtain its general $N$-soliton solution 
in 
simple explicit 
form. 
We also discuss how to obtain 
exact solutions 
such as 
the $N$-soliton solution 
of the vector mKdV equation (\ref{vmKdV1}); 
the obtained 
$N$-soliton forumula 
is 
different from 
the multisoliton formula 
proposed 
by Iwao and Hirota~\cite{Iwao1997}  
using the Hirota 
bilinear method~\cite{Hirota04}, 
and our formula 
has its own advantages. 
Section 4 is devoted to concluding remarks.

\section{Lax 
representations}

We start with 
the 
matrix 
generalization
of the 
nonreduced 
NLS system~\cite{AKNS73,AKNS74} 
proposed by Zakharov and Shabat 
as early as 1974~\cite{ZS74}: 
%
%
\begin{equation} 
\label{mNLS}
\left\{ 
\begin{split}
& \mathrm{i} Q_{t} + Q_{xx} - 2Q R Q = O, 
\\[0.5mm]
& \mathrm{i} R_{t} - R_{xx} + 2R Q R = O. 
\end{split} 
\right. 
\end{equation}
Here, 
$Q$ and $R$ are \mbox{$l_1 \times l_2$} 
and \mbox{$l_2 \times l_1$} 
(generally rectangular) 
matrices. 
Some 
relevant 
information and 
references on 
the 
matrix NLS system 
(\ref{mNLS})
can be found in~\cite{DM2010}. 

The Lax representation~\cite{Lax} for 
the matrix NLS system (\ref{mNLS})
is 
given by the 
following overdetermined 
linear system~\cite{Zakh,Konop1}: 
\begin{align}
& \left[
\begin{array}{c}
 \Psi_1  \\
 \Psi_2 \\
\end{array}
\right]_x 
= \left[
\begin{array}{cc}
-\mathrm{i}\zeta I_{l_1} & Q \\
 R & \mathrm{i}\zeta I_{l_2} \\
\end{array}
\right] 
\left[
\begin{array}{c}
 \Psi_1  \\
 \Psi_2 \\
\end{array}
\right],
\label{NLS-U}
\\[1.5mm]
& \left[
\begin{array}{c}
 \Psi_1  \\
 \Psi_2 \\
\end{array}
\right]_{t} 
= \left[
\begin{array}{cc}
-2\mathrm{i}\zeta^2 I_{l_1} -\mathrm{i} QR & 2 \zeta Q + \mathrm{i} Q_x \\
 2 \zeta R - \mathrm{i} R_x & 2\mathrm{i}\zeta^2 I_{l_2} +\mathrm{i} RQ \\
\end{array}
\right]
\left[
\begin{array}{c}
 \Psi_1  \\
 \Psi_2 \\
\end{array}
\right]. 
\label{NLS-V}
\end{align}
%
Here, 
$\zeta$ is 
a 
spectral parameter 
independent of $x$ and $t$, 
and 
$I_{l_1}$ and $I_{l_2}$ 
are the \mbox{$l_1 \times l_1 $} and \mbox{$l_2 \times l_2$} 
identity 
matrices, respectively; 
in this paper, 
we usually 
consider the \mbox{$l_1 = l_2$} case 
and 
omit 
the index 
of 
the 
identity 
matrix. 
%
%

The matrix NLS system (\ref{mNLS})
is 
a positive 
flow 
in 
the integrable hierarchy 
associated with 
the 
spectral problem (\ref{NLS-U}). 
The 
next higher flow 
in 
the 
integrable 
hierarchy 
is a matrix 
analog~\cite{Zakh,Konop1} 
of the nonreduced complex mKdV 
equation~\cite{AKNS73,AKNS74,Hirota73JMP}, 
i.e.
\begin{equation} 
\label{mmKdV}
\left\{ 
\begin{split}
& Q_{y} + Q_{xxx} - 3Q_x R Q -3 QRQ_x = O, 
\\[0.5mm]
& R_{y} + R_{xxx} -3 R_x Q R  -3 RQR_x = O. 
\end{split}
\right.
\end{equation}

To 
reduce the matrix NLS system (\ref{mNLS}) to 
the Kulish--Sklyanin model (\ref{rKS}) 
or, more generally, the matrix NLS hierarchy 
to the Kulish--Sklyanin hierarchy, 
we 
introduce 
\mbox{$2^{m-1} \times 2^{m-1}$} 
skew-Hermitian 
matrices \mbox{$\{ e_1, e_2, 
\ldots, e_{2m-1} \}$} 
that 
satisfy 
the anticommutation relations (\ref{a-commute0}). 
Then, we set 
\begin{equation}
\label{reduct}
Q =
q_1 I + \sum_{j=1}^{2m-1} q_{j+1} e_j, 
\hspace{5mm}
R = 
r_1 I - \sum_{j=1}^{2m-1} r_{j+1} e_j. 
\end{equation}
The matrices \mbox{$\{ I, e_1, e_2, \ldots, e_{2m-1} \}$} 
are assumed to be linearly independent. 
Lax representations involving the generators 
of the Clifford algebra 
(or quaternions 
in the 
\mbox{$m=2$} case) 
can be traced back to 
the references~\cite{PR79,EP79,KuSk81}.

As a natural 
extension 
of 
the complex 
conjugate, 
we 
define 
``Clifford 
conjugate"
denoted 
as 
\mbox{$\widehat{\;\;}$}, 
which 
acts on 
the linear span of 
\mbox{$\{ I, e_1, e_2, \ldots, e_{2m-1} \}$}
to reverse 
the sign 
of 
the coefficients 
of \mbox{$\{e_1, e_2, \ldots, e_{2m-1} \}$}. 
For instance, 
\[
\widehat{Q} = q_{1} I - \sum_{j=1}^{2m-1} q_{j+1} e_j,
\hspace{5mm}
\widehat{R} = r_{1} I + \sum_{j=1}^{2m-1} r_{j+1} e_j.
\]
Note that \mbox{$\widehat{\widehat{Q}
}= Q$}.  
Because 
of the anticommutation relations (\ref{a-commute0}), 
we have 
useful 
relations such as 
\begin{align}
& Q \hspace{1pt} \widehat{Q} = \widehat{Q} \hspace{1pt} Q 
 = \sca{\vt{q}}{\vt{q}} I, 
\nonumber 
\\[1mm]
& Q R + \widehat{R} \hspace{1pt}\widehat{Q} = \widehat{Q} \widehat{R} + R Q 
 =  2 \hspace{1pt} \sca{\vt{q}}{\vt{r}} I,
\nonumber 
\\[1mm]
& QRQ =
\left( QR+ \widehat{R} \hspace{1pt}\widehat{Q} \right) Q
- \widehat{R} \hspace{1pt}\widehat{Q} Q
= 2 \hspace{1pt} \sca{\vt{q}}{\vt{r}} Q
  - \sca{\vt{q}}{\vt{q}} \widehat{R},
\label{QRQ}
\\[1mm]
& RQR = R \left( QR+ \widehat{R} \hspace{1pt}\widehat{Q} \right) 
 - R \widehat{R} \hspace{1pt}\widehat{Q} 
= 2 \hspace{1pt} \sca{\vt{q}}{\vt{r}} R
  - \sca{\vt{r}}{\vt{r}} \widehat{Q},
\label{RQR}
\\[1mm]
& \left( I - \mu Q R \right) \left( I - \mu \widehat{R} \hspace{1pt} \widehat{Q} \right) 
= \left( 1- 2\mu \sca{\vt{q}}{\vt{r}} 
 + \mu^2  \sca{\vt{q}}{\vt{q}} \sca{\vt{r}}{\vt{r}} \right) I. 
\label{I-QR}
\end{align}
Here, 
\mbox{$\vt{q} = \left( q_1, q_2, \ldots, q_{2m} \right)$} 
and \mbox{$\vt{r} = \left( r_1, r_2, \ldots, r_{2m} \right)$} 
are 
$2m$-component 
row 
vectors; 
\mbox{$\sca{\,\cdot\,}{\,\cdot\,}$} 
denotes the 
standard 
scalar product, 
{\em e.g.}, 
\mbox{$\sca{\vt{q}}{\vt{r}} = \sum_{j=1}^{2m} q_j r_j$}, etc. 

Owing to 
(\ref{QRQ}) and (\ref{RQR}), 
the reduction (\ref{reduct}) simplifies 
the matrix NLS system (\ref{mNLS}) to the nonreduced 
Kulish--Sklyanin model: 
\begin{equation} 
\label{nonreduceKS}
\left\{ 
\begin{split}
& \mathrm{i} \vt{q}_t + \vt{q}_{xx} - 4 \sca{\vt{q}}{\vt{r}} \vt{q}
	+ 2 \sca{\vt{q}}{\vt{q}} \vt{r} = \vt{0},
\\[0.5mm]
& \mathrm{i} \vt{r}_t - \vt{r}_{xx} + 4 \sca{\vt{q}}{\vt{r}} \vt{r}
	- 2 \sca{\vt{r}}{\vt{r}} \vt{q} = \vt{0}. 
\end{split} 
\right. 
\end{equation}
Note that 
\mbox{$\sca{\vt{q}}{\vt{r}}$} 
and \mbox{$q_j r_k -q_k r_j$} are conserved densities for (\ref{nonreduceKS}).
By further imposing 
the general 
complex conjugation reduction 
\[
r_j = \sigma_j q_j^\ast, 
\hspace{5mm}
\sigma_j= \pm1, 
\hspace{5mm} j=1,2, \ldots, 2m, 
\]
we obtain the Kulish--Sklyanin model with a mixed focusing-defocusing 
nonlinearity: 
\begin{equation}
\mathrm{i} \vt{q}_t + \vt{q}_{xx} - 4 \sca{\vt{q}}{\vt{q}^\ast\Sigma} \vt{q}
	+ 2 \sca{\vt{q}}{\vt{q}} \vt{q}^\ast \Sigma = \vt{0}. 
\label{mixedKS}
\end{equation}
Here, 
\mbox{$\Sigma := \mathrm{diag} (\sigma_1, \sigma_2, \ldots, \sigma_{2m})$} 
is a diagonal matrix with 
each 
entry 
$\sigma_j$ 
equal to 
$+1$ or $-1$. 
In the following, we 
mainly 
consider the Kulish--Sklyanin model 
in the self-focusing case: 
\begin{equation}
\mathrm{i} \vt{q}_t + \vt{q}_{xx} + 4 \sca{\vt{q}}{\vt{q}^\ast} \vt{q}
	- 2 \sca{\vt{q}}{\vt{q}} \vt{q}^\ast = \vt{0}. 
\label{rKS2}
\end{equation}
The third-order symmetry 
of the nonreduced 
Kulish--Sklyanin model (\ref{nonreduceKS}) 
is obtained 
by imposing the reduction (\ref{reduct}) 
on the matrix 
complex mKdV 
system (\ref{mmKdV}) 
and 
noting 
the identities 
\mbox{$Q_x R Q + QRQ_x = \left( QRQ \right)_x - QR_x Q$}, 
\mbox{$R_x Q R + RQR_x = \left( RQR \right)_x - RQ_x R$} 
in view of (\ref{QRQ}) and (\ref{RQR}); 
by further 
setting \mbox{$\vt{r} = -\vt{q}$} 
(and thus \mbox{$R=-\widehat{Q}$} in (\ref{reduct})), 
(\ref{mmKdV}) 
reduces to 
the vector mKdV equation~\cite{Svi93,SviSok94}: 
\begin{align}
& \vt{q}_{y} 
+ 
\vt{q}_{xxx} + 6 \sca{\vt{q}}{\vt{q}}\vt{q}_x 
= \vt{0}. 
\label{vmKdV2}
\end{align}

The matrix NLS hierarchy 
can be solved using the inverse scattering method 
based on 
the 
spectral 
problem (\ref{NLS-U}), 
so 
the exact solutions such as the $N$-soliton solution 
of the 
matrix NLS system (\ref{mNLS}), 
as well as 
the third-order symmetry (\ref{mmKdV}), 
can be obtained explicitly in closed form. 
Thus, 
the exact solutions 
of the Kulish--Sklyanin model (\ref{rKS2}),  
as well as the vector mKdV equation (\ref{vmKdV2}), 
can
also be obtained by 
imposing 
the corresponding reduction conditions 
on the scattering data involved in the solution. 
However, this 
approach is useful only 
if the number of components 
or solitons is small enough. 
Indeed, 
the obtained 
formula for 
the $N$-soliton solution of the $2m$-component 
Kulish--Sklyanin model (\ref{rKS2}) 
involves the inverse of an \mbox{$N \times N$} block matrix, 
where 
each block is a \mbox{$2^{m-1} \times 2^{m-1}$} matrix 
taking values in the linear span of 
\mbox{$\{ I, e_1, e_2, \ldots, e_{2m-1} \}$}. 
The formula 
is 
too bulky and 
not 
a mathematically 
tractable 
object 
for 
\mbox{$2m > 4$} and \mbox{$N 
> 2$}. 

In the four-component 
case 
\mbox{$(2m=4)$}, 
the reduction 
(\ref{reduct}) is no longer a restriction. 
Indeed, 
one can 
employ \mbox{$2 \times 2$} 
Pauli's 
matrices 
multiplied by the imaginary unit 
$\mathrm{i}$ 
as a matrix representation for 
\mbox{$\{e_1, e_2, e_3 \}$}: 
\[
e_1 = \left[
\begin{array}{cc}
 0 &  \mathrm{i}\\
 \mathrm{i} & 0 \\
\end{array}
\right], 
\hspace{5mm}
e_2 = \left[
\begin{array}{cc}
 0 & 1 \\
 -1 & 0 \\
\end{array}
\right], 
\hspace{5mm}
e_3 = \left[
\begin{array}{cc}
 \mathrm{i} & 0 \\
 0 & -\mathrm{i} \\
\end{array}
\right]. 
\]
These 
matrices together with the identity matrix 
form 
a basis, 
i.e., 
any \mbox{$2 \times 2$} complex matrix can be expressed 
as 
a linear combination of \mbox{$\{I, e_1, e_2, e_3 \}$}; 
thus, (\ref{reduct}) is merely a linear transformation 
mixing the elements 
in the 
\mbox{$2 \times 2$} 
matrices 
$Q$ and $R$. 
In the self-focusing case, this 
linear transformation 
reads 
\[
Q= \left[
\begin{array}{cc}
 q_1 + \mathrm{i} q_4 &  \mathrm{i} q_2 +q_3 \\
 \mathrm{i} q_2 - q_3 & q_1 -\mathrm{i} q_4 \\
\end{array}
\right], 
\hspace{5mm}
R= -Q^\dagger = 
\left[
\begin{array}{cc}
 -q_1^\ast + \mathrm{i} q_4^\ast &  \mathrm{i} q_2^\ast +q_3^\ast \\
 \mathrm{i} q_2^\ast - q_3^\ast & -q_1^\ast -\mathrm{i} q_4^\ast \\
\end{array}
\right], 
\]
where $Q$ satisfies 
the matrix NLS equation (\ref{redNLS}). 
Clearly, the $N$-soliton solution of the 
Kulish--Sklyanin model (\ref{rKS2}) 
for a four-component 
vector $\vt{q}$ 
can be directly 
obtained from the $N$-soliton solution of 
the 
matrix NLS equation (\ref{redNLS}) 
for a \mbox{$2 \times 2$} 
matrix $Q$ 
by applying 
this linear transformation. 

The Kulish--Sklyanin model (\ref{rKS2}) 
for a three-component vector $\vt{q}$ 
is 
obtained by setting one 
component, say $q_3$, 
in the four-component case as 
identically 
zero. 
The reduction 
\mbox{$q_3=0$} corresponds to 
the restriction of $Q$ 
to a 
symmetric matrix~\cite{ForKul83}; 
the corresponding reduction of the $N$-soliton solution 
from the four-component case to the three-component case 
is straightforward~\cite{Ieda04-1,Ieda04-2}.

It is clear by setting \mbox{$q_2=q_3=0$} in the above representation 
that the Kulish--Sklyanin model (\ref{rKS2}) 
in the two-component case, say \mbox{$\vt{q}=\left( q_1, q_4 \right)$} 
can be decoupled into two scalar NLS equations 
in the variables \mbox{$q_1 \pm \mathrm{i} q_4$}~\cite{ParkShin99}. 
Thus, any solution of the two-component Kulish--Sklyanin model 
can be written as a linear combination 
of 
two 
solutions of the scalar NLS equation; 
in this sense, the two-component case 
is trivial and less interesting. 
The rank-$1$ one-soliton solution in the two-component case 
is 
\begin{equation}
\vt{q} (x,t) = 2 \eta 
\sech \left[ 2 \eta (x + 4 \xi t) + \alpha \right] 
 \mathrm{e}^{-2 \mathrm{i}\xi x - 4 \mathrm{i} (\xi^2 - \eta^2) t +\mathrm{i}\varphi} 
 \left( \, \frac{1}{2},\; \pm \frac{\mathrm{i}}{2} \, \right),
\nonumber
\end{equation}
and the rank-$2$ one-soliton solution 
is 
\begin{align}
\vt{q} (x,t) &= 2 \eta 
\sech \left[ 2 \eta (x + 4 \xi t) + \alpha_1 \right] 
 \mathrm{e}^{-2 \mathrm{i}\xi x - 4 \mathrm{i} (\xi^2 - \eta^2) t +\mathrm{i}\varphi_1} 
 \left( \, \frac{1}{2}, \; -\frac{\mathrm{i}}{2} \, \right)
\nonumber \\ 
& \hphantom{=} \; \, \mbox{} + 2 \eta 
\sech \left[ 2 \eta (x + 4 \xi t) + \alpha_2 \right] 
 \mathrm{e}^{-2 \mathrm{i}\xi x - 4 \mathrm{i} (\xi^2 - \eta^2) t +\mathrm{i}\varphi_2}
  \left( \, \frac{1}{2}, \; \frac{\mathrm{i}}{2} \, \right).
\nonumber
\end{align}
Here, \mbox{$\eta > 0$}
and the other parameters are 
real constants. 
This implies that 
the rank-$1$ one-soliton solution 
in the general component case 
is 
\begin{equation}
\vt{q} (x,t) = 2 \eta 
\sech \left[ 2 \eta (x + 4 \xi t) + \alpha \right] 
 \mathrm{e}^{-2 \mathrm{i}\xi x - 4 \mathrm{i} (\xi^2 - \eta^2) t} \vt{u}
\label{rank1}
\end{equation}
where \mbox{$\sca{\vt{u}}{\vt{u}}=0$} and \mbox{$\sca{\vt{u}}{\vt{u}^\ast}= \frac{1}{2}
$}, 
and the rank-$2$ one-soliton solution is 
\begin{align}
\vt{q} (x,t) &= 2 \eta 
\sech \left[ 2 \eta (x + 4 \xi t) + \alpha_1 \right] 
 \mathrm{e}^{-2 \mathrm{i}\xi x - 4 \mathrm{i} (\xi^2 - \eta^2) t +\mathrm{i}\varphi} 
\vt{u}
\nonumber \\ 
& \hphantom{=} \; \, \mbox{} + 2 \eta 
\sech \left[ 2 \eta (x + 4 \xi t) + \alpha_2 \right] 
 \mathrm{e}^{-2 \mathrm{i}\xi x - 4 \mathrm{i} (\xi^2 - \eta^2) t +\mathrm{i}\varphi} 
\vt{u}^\ast
\label{rank2}
\end{align}
where \mbox{$\sca{\vt{u}}{\vt{u}}=0$} 
and 
\mbox{$\sca{\vt{u}}{\vt{u}^\ast}= \frac{1}{2}
$}. 

In the two-component case, 
the Kulish--Sklyanin model 
with a mixed focusing-defocusing nonlinearity 
is more interesting than the model with a simple 
focusing (or defocusing) nonlinearity.  
Indeed, 
(\ref{mixedKS}) 
with \mbox{$\vt{q}=\left( q_1, q_4 \right)$} 
and \mbox{$\Sigma = \mathrm{diag} (-1, 1)$}~\cite{Ozer98}: 
\begin{equation} 
\label{2comKS}
\left\{ 
\begin{split}
& \mathrm{i} q_{1,t} + q_{1,xx} + 2 \left( |q_1|^2 -2 |q_4|^2 \right) q_1
	- 2 q_4^2 q_1^\ast = 0,
\\[0.5mm]
& \mathrm{i} q_{4,t} + q_{4,xx} + 2 \left( 2 |q_1|^2 - |q_4|^2 \right) q_4
	+ 2 q_1^2 q_4^\ast = 0, 
\end{split} 
\right. 
\end{equation}
is obtained from the matrix NLS system (\ref{mNLS})
through the reduction 
\[
Q= \left[
\begin{array}{cc}
 q_1 + \mathrm{i} q_4 & 0 \\
 0 & q_1 -\mathrm{i} q_4 \\
\end{array}
\right], 
\hspace{5mm}
R= 
\left[
\begin{array}{cc}
 -q_1^\ast - \mathrm{i} q_4^\ast & 0 \\
 0 & -q_1^\ast + \mathrm{i} q_4^\ast \\
\end{array}
\right]. 
\]
Thus, 
the two-component Kulish--Sklyanin model 
with a mixed focusing-defocusing nonlinearity (\ref{2comKS}) 
is equivalent to the nonreduced scalar NLS system~\cite{AKNS73,AKNS74}: 
\begin{equation} 
\nonumber 
\left\{ 
\begin{split}
& \mathrm{i} q_{t} + q_{xx} - 2q^2 r = 0, 
\\
& \mathrm{i} r_{t} - r_{xx} + 2r^2 q = 0, 
\end{split} 
\right. 
\end{equation} 
through the 
linear change of variables 
\mbox{$q = q_1 + \mathrm{i} q_4$}, 
\mbox{$r = -q_1^\ast - \mathrm{i} q_4^\ast $} 
(or \mbox{$q = q_1 - \mathrm{i} q_4$}, 
\mbox{$r = -q_1^\ast + \mathrm{i} q_4^\ast $}). 

In this paper, we aim to obtain 
a 
compact
and 
tractable 
expression for 
the 
$N$-soliton solution
of the Kulish--Sklyanin model (\ref{rKS2}), 
which 
is 
valid 
for 
an arbitrary number of components 
and does not involve the generators of the Clifford algebra. 
To 
derive such a ``classical" expression, 
we first rewrite the spectral problem (\ref{NLS-U}) 
under the reduction (\ref{reduct}) to a more convenient 
form. 
We consider 
a linear eigenfunction 
the first 
component 
of which 
is an invertible 
matrix; 
then, 
the spectral problem (\ref{NLS-U}) can be 
rewritten in terms of 
\mbox{$P := \Psi_2 \Psi_1^{-1}$} 
as a 
matrix Riccati 
equation 
(see~\cite{Chen1,
WSK,KW75} for the scalar case and~\cite{Haberman,Pri81} for the vector case):
%
\begin{equation}
P_x = R + 2 \mathrm{i} \zeta P -P Q P.
\label{NLS-R1}
\end{equation}
Thus, 
under the reduction (\ref{reduct}) 
and appropriate 
boundary conditions, 
we can 
confine 
$P$
to the linear span of 
\mbox{$\{ I, e_1, e_2, \ldots, e_{2m-1} \}$}. 
By setting 
\begin{equation}
P = p_1 I - \sum_{j=1}^{2m-1} p_{j+1} e_j, 
\hspace{5mm} 
\vt{p} = \left( p_1, p_2, \ldots, p_{2m} \right)
\nonumber 
\end{equation}
and noting the relation (\ref{QRQ}), 
we can 
simplify 
(\ref{NLS-R1}) 
to a vector Riccati 
equation: 
\begin{equation}
\vt{p}_x = \vt{r} + 2 \mathrm{i} \zeta \vt{p} 
 -2 \sca{\vt{p}}{\vt{q}}\vt{p} + \sca{\vt{p}}{\vt{p}}\vt{q}. 
\label{vRiccati}
\end{equation}
We introduce the scalar 
denominator $f$ 
and the vector numerator $\vt{g}$ as 
\begin{subequations}
\label{p_linear}
\begin{equation}
\vt{p} = \frac{\vt{g}}{f}, 
\end{equation}
and set 
\begin{equation}
\sca{\vt{g}}{\vt{g}} = f \hspace{1pt} h. 
\label{ggfh}
\end{equation}
\end{subequations}
Noting 
the 
freedom to multiply $f$ and $\vt{g}$ by any common factor, 
we can linearize 
the vector Riccati equation (\ref{vRiccati}) 
as 
\[
\left[
\begin{array}{c}
 f  \\
 \vt{g}^T \\
 h \\
\end{array}
\right]_x 
= \left[
\begin{array}{ccc}
-2\mathrm{i}\zeta & 2 \vt{q} & 0 \\
 \vt{r}^T & O & \vt{q}^T \\
 0 & 2 \vt{r} & 2\mathrm{i}\zeta \\
\end{array}
\right] 
\left[
\begin{array}{c}
 f  \\
 \vt{g}^T \\
 h \\
\end{array}
\right], 
\]
where the superscript ${}^T$ denotes the matrix transpose. 
This 
spectral problem 
is the spatial part of a
nonstandard Lax representation for 
the nonreduced Kulish--Sklyanin model (\ref{nonreduceKS})~\cite{Adler94}; 
this kind of nonstandard 
spectral problem 
first 
appeared 
in~\cite{AKNS74,Ab78} 
through the investigation 
of the squared 
eigenfunctions 
associated with the scalar NLS hierarchy
and 
a certain vector generalization was 
studied in~\cite{Eich80,Eich81}. 
The corresponding 
time part of the Lax representation 
can, in principle, 
be derived from (\ref{NLS-V}) in 
an analogous manner, 
but it is easier to 
obtain 
the temporal Lax matrix 
from the compatibility condition 
as a truncated power series 
in 
the spectral parameter $\zeta$~\cite{AKNS73,AKNS74,Ab78}. 
For later convenience, we rescale $\vt{q}$, $\vt{r}$ 
and $\vt{g}^T$ by a factor of $1/\sqrt{2}$
and set \mbox{$2\zeta=: \lambda$} 
to reformulate the nonstandard Lax representation 
in a more symmetric and concise 
form. 

\begin{proposition}
The nonreduced Kulish--Sklyanin model 
with an arbitrary number of vector 
components: 
\begin{equation} 
\label{nonreduceKS2}
\left\{ 
\begin{split}
& \mathrm{i} \vt{q}_t + \vt{q}_{xx} - 2 \sca{\vt{q}}{\vt{r}} \vt{q}
	+ \sca{\vt{q}}{\vt{q}} \vt{r} = \vt{0},
\\[0.5mm]
& \mathrm{i} \vt{r}_t - \vt{r}_{xx} + 2 \sca{\vt{q}}{\vt{r}} \vt{r}
	- \sca{\vt{r}}{\vt{r}} \vt{q} = \vt{0}, 
\end{split} 
\right. 
\end{equation}
is 
equivalent to 
the compatibility condition for 
the 
overdetermined linear system~\cite{Adler94}: 
\begin{align}
& \left[
\begin{array}{c}
 \psi_1  \\
 \vt{\psi}_2 \\
 \psi_3 \\
\end{array}
\right]_x 
= \left[
\begin{array}{ccc}
-\mathrm{i}\lambda &  \vt{q} & 0 \\
 \vt{r}^T & O & \vt{q}^T \\
 0 & \vt{r} & \mathrm{i}\lambda \\
\end{array}
\right] 
\left[
\begin{array}{c}
 \psi_1  \\
 \vt{\psi}_2 \\
 \psi_3 \\
\end{array}
\right], 
\label{Jordan_U}
\\[2mm]
& \left[
\begin{array}{c}
 \psi_1  \\
 \vt{\psi}_2 \\
 \psi_3 \\
\end{array}
\right]_t 
= \left[
\begin{array}{ccc}
-\mathrm{i}\lambda^2 -\mathrm{i} \sca{\vt{q}}{\vt{r}} 
	&  \lambda \vt{q} +\mathrm{i}\vt{q}_x & 0 \\
 \lambda \vt{r}^T - \mathrm{i} \vt{r}^T_x 
	& \mathrm{i} \vt{r}^T \vt{q} - \mathrm{i} \vt{q}^T \vt{r} 
	& \lambda \vt{q}^T + \mathrm{i} \vt{q}^T_x \\
 0 &  \lambda \vt{r}-\mathrm{i} \vt{r}_x 
	& \mathrm{i} \lambda^2 + \mathrm{i} \sca{\vt{q}}{\vt{r}}\\
\end{array}
\right] 
\left[
\begin{array}{c}
 \psi_1  \\
 \vt{\psi}_2 \\
 \psi_3 \\
\end{array}
\right].
\label{Jordan_V}
\end{align}
Here, $\lambda$ is a constant spectral parameter, 
$\vt{q}$ and $\vt{r}$ are row vectors and $\vt{\psi}_2$ 
is a column vector. 
\end{proposition}

By 
rewriting 
the spectral problem 
(\ref{Jordan_U}) 
as the adjoint problem 
\begin{equation}
\left[
\begin{array}{ccc}
 \! \psi_3 \! & \! -\vt{\psi}_2^T \! & \! \psi_1 \! 
\end{array}
\right]_x 
= -\left[
\begin{array}{ccc}
 \! \psi_3 \! & \! -\vt{\psi}_2^T \! & \! \psi_1 \! 
\end{array}
\right] \left[
\begin{array}{ccc}
-\mathrm{i}\lambda &  \vt{q} & 0 \\
 \vt{r}^T & O & \vt{q}^T \\
 0 & \vt{r} & \mathrm{i}\lambda \\
\end{array}
\right], 
\label{adj}
\end{equation}
or noting the 
identity 
\begin{equation}
\left[
\begin{array}{ccc}
 \! \psi_3 \! & \! -\vt{\psi}_2^T \! & \! \psi_1 \! 
\end{array}
\right]
\left[
\begin{array}{c}
 \psi_1  \\
 \vt{\psi}_2 \\
 \psi_3 \\
\end{array}
\right]_x 
= \left[
\begin{array}{ccc}
\! \psi_3 \! & \! -\vt{\psi}_2^T \! & \! \psi_1 \!
\end{array}
\right]
\left[
\begin{array}{ccc}
-\mathrm{i}\lambda &  \vt{q} & 0 \\
 \vt{r}^T & O & \vt{q}^T \\
 0 & \vt{r} & \mathrm{i}\lambda \\
\end{array}
\right] 
\left[
\begin{array}{c}
 \psi_1  \\
 \vt{\psi}_2 \\
 \psi_3 \\
\end{array}
\right] =0,
\nonumber 
\end{equation}
and similar for 
$t$-differentiation, 
we notice 
that 
the quantity 
\mbox{2$\psi_1\psi_3 - 
\sca{\vt{\psi}_2}{\vt{\psi}_2}$}
is 
a constant. 
In fact, the derivation from the standard Lax representation, 
(\ref{NLS-U}) and (\ref{NLS-V}), through the reduction (\ref{reduct}) 
implies that 
we 
only need to consider 
linear eigenfunctions 
satisfying the condition 
\mbox{$2\psi_1\psi_3 = 
\sca{\vt{\psi}_2}{\vt{\psi}_2}$} (cf.~(\ref{ggfh})). 
In this paper, we use the notation 
\mbox{$\sca{\,\cdot\,}{\,\cdot\,}$}
to denote 
the scalar product of two row vectors 
as well as 
column vectors. 

%

\begin{proposition}
The third-order symmetry of the nonreduced Kulish--Sklyanin model 
\mbox{$(\ref{nonreduceKS2})$} 
reads~\cite{Ad,Svi92}
\begin{equation} 
\label{nonreduceKS3}
\left\{ 
\begin{split}
& \vt{q}_y + \vt{q}_{xxx} -3 \sca{\vt{q}_x}{\vt{r}} \vt{q}
	-3 \sca{\vt{q}}{\vt{r}} \vt{q}_x
	+3 \sca{\vt{q}}{\vt{q}_x} \vt{r} = \vt{0},
\\[0.5mm]
& \vt{r}_y + \vt{r}_{xxx} -3 \sca{\vt{q}}{\vt{r}_x} \vt{r}
	-3 \sca{\vt{q}}{\vt{r}} \vt{r}_x
	+3 \sca{\vt{r}}{\vt{r}_x} \vt{q} = \vt{0}. 
\end{split} 
\right. 
\end{equation}
The reduction \mbox{$\vt{r}=-\vt{q}$} simplifies 
\mbox{$(\ref{nonreduceKS3})$} 
to the vector mKdV equation~\cite{Svi93,SviSok94}: 
\begin{equation}
\label{vmKdV3}
\vt{q}_y + \vt{q}_{xxx} +3 \sca{\vt{q}}{\vt{q}} \vt{q}_x = \vt{0},  
\end{equation}
which 
is obtained as the compatibility condition for 
the 
overdetermined linear system: 
\begin{align}
& \left[
\begin{array}{c}
 \psi_1  \\
 \vt{\psi}_2 \\
 \psi_3 \\
\end{array}
\right]_x 
= \left[
\begin{array}{ccc}
-\mathrm{i}\lambda &  \vt{q} & 0 \\
 -\vt{q}^T & O & \vt{q}^T \\
 0 & -\vt{q} & \mathrm{i} \lambda \\
\end{array}
\right] 
\left[
\begin{array}{c}
 \psi_1  \\
 \vt{\psi}_2 \\
 \psi_3 \\
\end{array}
\right], 
\label{Jordan_U2}
\\[2mm]
& \left[
\begin{array}{c}
 \psi_1  \\
 \vt{\psi}_2 \\
 \psi_3 \\
\end{array}
\right]_y
= \left[
\begin{array}{ccc}
-\mathrm{i}\lambda^3 +\mathrm{i} \lambda \sca{\vt{q}}{\vt{q}} 
	& \lambda^2 \vt{q} +\mathrm{i} \lambda \vt{q}_x -\vt{\alpha} & 0 \\
 -\lambda^2 \vt{q}^T + \mathrm{i}\lambda \vt{q}^T_x +\vt{\alpha}^T 
	& -2\vt{q}^T_x \vt{q} + 2\vt{q}^T \vt{q}_x 
	& \lambda^2 \vt{q}^T + \mathrm{i} \lambda \vt{q}^T_x - \vt{\alpha}^T \\
 0 & -\lambda^2 \vt{q} + \mathrm{i} \lambda \vt{q}_x +\vt{\alpha} 
	& \mathrm{i}\lambda^3 - \mathrm{i} \lambda \sca{\vt{q}}{\vt{q}}\\
\end{array}
\right] 
\left[
\begin{array}{c}
 \psi_1  \\
 \vt{\psi}_2 \\
 \psi_3 \\
\end{array}
\right],
\label{Jordan_V2}
\end{align}
with 
\mbox{$\vt{\alpha}:=\vt{q}_{xx}+\sca{\vt{q}}{\vt{q}}\vt{q}$}.
\end{proposition}


\section{Darboux 
transformations 
and 
multisoliton 
solutions}

\subsection{Darboux transformations}

We 
propose 
the binary 
Darboux
(or Zakharov--Shabat
dressing) 
transformation~\cite{ZS79,Sall82,Chud2,ZaMi80,MatSall91}  
that can be applied 
to the spectral problem (\ref{Jordan_U}) 
associated with the 
Kulish--Sklyanin hierarchy. 
This can be obtained 
by 
considering how 
the binary Darboux transformation
for  
the spectral problem (\ref{NLS-U}) 
acts on 
\mbox{$P = \Psi_2 \Psi_1^{-1}$} 
under the reduction (\ref{reduct}), 
confining the result 
to the linear span of 
\mbox{$\{ I, e_1, e_2, \ldots, e_{2m-1} \}$} 
and then 
linearizing 
the discrete 
vector equation 
through the transformation 
(\ref{p_linear}). 

Let $\Lambda$ be 
the block 
anti-diagonal matrix: 
\[
\Lambda := 
\left[
\begin{array}{ccc}
 & & 1 \\
 & -I & \\
 1 & & \\
\end{array}
\right], 
\hspace{5mm} \Lambda^T = \Lambda, 
\hspace{5mm} \Lambda^2 =I,  
\]
and denote a 
column-vector 
eigenfunction of the 
spectral problem (\ref{Jordan_U}) at \mbox{$\lambda=\mu$} 
and its matrix transpose 
(i.e., row vector) as 
\[
\Ket{\mu} := 
\left. 
\left[ 
\begin{array}{c}
 \psi_1  \\
 \vt{\psi}_2 \\
 \psi_3 \\
\end{array}
\right] \right|_{\lambda=\mu}, 
\hspace{5mm} \Bra{\mu} := 
\left. \left[
\begin{array}{ccc}
 \! \psi_1 \! & \! \vt{\psi}_2^T \! & \! \psi_3 \! 
\end{array}
\right] \right|_{\lambda=\mu}, 
\]
which 
satisfy the condition 
\[
\Braket{\mu|\Lambda|\mu} 
= \left. 2 \psi_1 \psi_3 -\sca{\vt{\psi}_2}{\vt{\psi}_2} \right|_{\lambda=\mu} =0.
\]
In the same 
manner, 
we 
introduce 
a 
column-vector 
eigenfunction $\Ket{\nu}$ of the 
spectral problem (\ref{Jordan_U}) at \mbox{$\lambda=\nu$}
and its matrix transpose $\Bra{\nu}$. 

\begin{proposition}
\label{prop3.1}
The spectral problem \mbox{$(\ref{Jordan_U})$} is form-invariant 
under the action of the binary Darboux transformation
defined as 
\begin{equation}
\left[
\begin{array}{c}
 \widetilde{\psi}_1  \\
 \widetilde{\vt{\psi}}_2 \\
 \widetilde{\psi}_3 \\
\end{array}
\right] 
\propto
\left\{ I + 
\left( \frac{\nu-\mu}{\lambda- \nu} \right)
	\frac{\Ket{\mu} \Bra{\nu}\Lambda}{\Braket{\nu|\Lambda|\mu}}
 + \left( \frac{\mu-\nu}{\lambda- \mu} \right)
	\frac{\Ket{\nu} \Bra{\mu}\Lambda}{\Braket{\mu|\Lambda|\nu}}
\right\}
\left[
\begin{array}{c}
 \psi_1  \\
 \vt{\psi}_2 \\
 \psi_3 \\
\end{array}
\right], 
\label{Dar_def}
\end{equation}
up to an overall constant, 
where 
\mbox{$\Braket{\mu|\Lambda|\mu} = \Braket{\nu|\Lambda|\nu} =0$}
and the 
transformed 
potentials $\widetilde{\vt{q}}$ and $\widetilde{\vt{r}}$ 
are 
given by 
\begin{subequations}
\label{qr_til}
\begin{align}
\widetilde{\vt{q}} & = \vt{q} +\mathrm{i} \left( \mu-\nu \right)
\frac{\left( \Ket{\mu}\Bra{\nu}-\Ket{\nu}\Bra{\mu} \right)_{12}}
	{\Braket{\nu|\Lambda|\mu}}, 
\\[2mm]
\widetilde{\vt{r}} &= \vt{r} +\mathrm{i} \left( \nu-\mu \right)
\frac{\left( \Ket{\mu}\Bra{\nu}-\Ket{\nu}\Bra{\mu} \right)_{32}}
	{\Braket{\nu|\Lambda|\mu}}.
\end{align}
\end{subequations}
Here, the subscripts 
${}_{12}$ and ${}_{32}$ denote 
the 
$(1,2)$ and $(3,2)$ 
sub-matrices (row vectors in this case) 
in 
the \mbox{$3 \times 3$} block matrix. 
\end{proposition}

In (\ref{Dar_def}), 
\[
	\frac{\Ket{\mu}}{\Braket{\nu|\Lambda|\mu}}, \hspace{5mm} 
	\frac{\Ket{\nu}}{\Braket{\mu|\Lambda|\nu}}
\]
provide 
linear eigenfunctions of the transformed spectral problem 
at \mbox{$\lambda=\nu$} and \mbox{$\lambda=\mu$}, respectively; 
for a suitable choice of $\Ket{\mu}$ and $\Ket{\nu}$, 
these correspond to 
bound states generated by the 
binary Darboux transformation. 
%
Note that 
overall factors of 
$\Ket{\mu}$ and $\Ket{\nu}$ 
play no role in 
the definition of 
the binary Darboux transformation. 

If 
$\Ket{\mu}$ and $\Ket{\nu}$ 
satisfy 
not only the spectral problem \mbox{$(\ref{Jordan_U})$} 
but also the isospectral 
evolution equation \mbox{$(\ref{Jordan_V})$}
at \mbox{$\lambda=\mu$} and \mbox{$\lambda=\nu$}, respectively, 
the binary Darboux transformation (\ref{Dar_def}) 
preserves the Lax representation,  
\mbox{$(\ref{Jordan_U})$} and \mbox{$(\ref{Jordan_V})$}, form-invariant
with 
the potentials 
transformed 
as 
\mbox{$\vt{q} \to \widetilde{\vt{q}}$} and 
\mbox{$\vt{r} \to \widetilde{\vt{r}}$}. 
This 
is also 
true 
for other flows of the 
integrable 
hierarchy. 
Thus, (\ref{qr_til}) can 
be used to 
generate a new nontrivial 
solution of the 
Kulish--Sklyanin 
hierarchy 
from 
its 
trivial 
solution. 

Similar 
results on the 
Darboux transformations have 
been 
obtained by Mikhailov and coworkers 
(see, in particular, the pioneering paper~\cite{ZaMi80}
and the recent papers~\cite{JPW01,JPW02}). 

The Darboux matrix defined in (\ref{Dar_def}): 
\begin{equation} 
D_{\mu,\nu} = I + 
\left( \frac{\nu-\mu}{\lambda- \nu} \right)
	\frac{\Ket{\mu} \Bra{\nu}\Lambda}{\Braket{\nu|\Lambda|\mu}}
 + \left( \frac{\mu-\nu}{\lambda- \mu} \right)
	\frac{\Ket{\nu} \Bra{\mu}\Lambda}{\Braket{\mu|\Lambda|\nu}}
\nonumber
\end{equation}
has the important invariance property: 
\begin{equation}
D_{\mu,\nu}^T \Lambda D_{\mu,\nu} = \Lambda, 
\nonumber 
\end{equation}
which 
implies \mbox{$\det D_{\mu,\nu}=1$} and 
\begin{equation}
 D_{\mu,\nu}^{-1} = \Lambda D_{\mu,\nu}^T \Lambda. 
\nonumber 
\end{equation}
Thus, 
the 
constant quantity 
\mbox{$2 \psi_1 \psi_3 -\sca{\vt{\psi}_2}{\vt{\psi}_2}$} 
for any linear eigenfunction 
is invariant under the binary Darboux transformation, i.e.\
\[
2 \widetilde{\psi}_1 \widetilde{\psi}_3 
	-\sca{\widetilde{\vt{\psi}}_2}{\widetilde{\vt{\psi}}_2} 
= 2 \psi_1 \psi_3 -\sca{\vt{\psi}_2}{\vt{\psi}_2}.
\]
In particular, if we start from a linear eigenfunction 
satisfying the condition 
\mbox{$2 \psi_1 \psi_3 =\sca{\vt{\psi}_2}{\vt{\psi}_2}$}, 
any linear eigenfunction 
generated by 
iterations of the binary Darboux transformation
also satisfies the same condition. 

We can 
consider 
an arbitrary number of 
iterations 
of the binary Darboux transformation 
(\ref{Dar_def}) with different values of $\mu$  
and $\nu$ in each step. 
For instance, 
the twofold binary Darboux transformation 
can be 
represented by the Darboux matrix: 
\begin{align} 
\widetilde{D}_{\mu_2,\nu_2} D_{\mu_1,\nu_1} &= 
 \left\{ I + 
 \left( \frac{\nu_2-\mu_2}{\lambda- \nu_2} \right)
	\frac{\Ket{\widetilde{\mu}_2} \Bra{\widetilde{\nu}_2}
	\Lambda}{\Braket{\widetilde{\nu}_2|\Lambda|\widetilde{\mu}_2}}
 + \left( \frac{\mu_2-\nu_2}{\lambda- \mu_2} \right)
	\frac{\Ket{\widetilde{\nu}_2} \Bra{\widetilde{\mu}_2}\Lambda}
	{\Braket{\widetilde{\mu}_2|\Lambda|\widetilde{\nu}_2}} \right\}
\nonumber \\[2mm]
& \hphantom{=} \;\, \mbox{} \times
 \left\{ I + 
 \left( \frac{\nu_1-\mu_1}{\lambda- \nu_1} \right)
	\frac{\Ket{\mu_1} \Bra{\nu_1}\Lambda}{\Braket{\nu_1|\Lambda|\mu_1}}
 + \left( \frac{\mu_1-\nu_1}{\lambda- \mu_1} \right)
	\frac{\Ket{\nu_1} \Bra{\mu_1}\Lambda}{\Braket{\mu_1|\Lambda|\nu_1}} \right\}, 
\nonumber 
\end{align}
where 
\begin{align} 
\Ket{\widetilde{\mu}_2} &= 
 \Ket{\mu_2} + \left( \frac{\nu_1-\mu_1}{\mu_2- \nu_1} \right)
	\frac{\Braket{\nu_1|\Lambda|\mu_2}}{\Braket{\nu_1|\Lambda|\mu_1}} \Ket{\mu_1}
 + \left( \frac{\mu_1-\nu_1}{\mu_2- \mu_1} \right)
	\frac{\Braket{\mu_1|\Lambda|\mu_2}}{\Braket{\mu_1|\Lambda|\nu_1}} \Ket{\nu_1}, 
\nonumber \\[2mm]
\Ket{\widetilde{\nu}_2} &= 
 \Ket{\nu_2} + \left( \frac{\nu_1-\mu_1}{\nu_2- \nu_1} \right)
	\frac{\Braket{\nu_1|\Lambda|\nu_2}}{\Braket{\nu_1|\Lambda|\mu_1}} \Ket{\mu_1}
 + \left( \frac{\mu_1-\nu_1}{\nu_2- \mu_1} \right)
	\frac{\Braket{\mu_1|\Lambda|\nu_2}}{\Braket{\mu_1|\Lambda|\nu_1}} \Ket{\nu_1}, 
\nonumber 
\end{align}
and \mbox{$\Braket{\mu_j|\Lambda|\mu_j} = \Braket{\nu_j|\Lambda|\nu_j} =0 
\;\, (j=1,2)$}.

Noting that 
a multifold binary 
Darboux 
transformation 
can be defined 
as 
the order-independent 
composition of 
binary Darboux 
transformations, 
%
we 
can assume 
that 
the $N$-fold binary Darboux transformation 
takes the following form (cf.~\cite{ZaMi80}): 
\begin{equation}
D_{\lambda_1, \lambda_2, \ldots, \lambda_{2N}} 
= I +\sum_{k=1}^{2N}  \frac{1}{\lambda-\lambda_k} \left( \sum_{j=1}^{2N} g_{jk} 
\Ket{\lambda_j} \right) \Bra{\lambda_k} \Lambda. 
\label{D_assump}
\end{equation}
Here, 
\mbox{$\{ \lambda_1, \lambda_2, \ldots, \lambda_{2N} \}$} 
are pairwise distinct constants, 
$g_{jk}$ is 
a scalar function 
to be determined,  
$\Ket{\lambda_j}$ is a nonzero 
column-vector eigenfunction of the spectral problem 
(\ref{Jordan_U}) at \mbox{$\lambda=\lambda_j$}, 
and $\Ket{\lambda_j}$ and its matrix transpose (i.e., row vector) 
$\Bra{\lambda_j}$ satisfy the condition 
\mbox{$\Braket{\lambda_j|\Lambda|\lambda_j} =0$}. 
Then, 
substituting (\ref{D_assump})
into 
the invariance property: 
\begin{equation}
D_{\lambda_1, \lambda_2, \ldots, \lambda_{2N}}^T \Lambda 
	D_{\lambda_1, \lambda_2, \ldots, \lambda_{2N}} = \Lambda, 
\nonumber 
\end{equation}
and noting that this is an identity in $\lambda$, we 
obtain the relations: 
\begin{align}
& \; g_{kk} = 0, 
\hspace{4mm} 
k=1,2, \ldots, 2N, 
\nonumber \\
& \left( \lambda_k - \lambda_j  \right) g_{jk} +\sum_{i=1}^{2N} \sum_{l=1}^{2N} 
g_{ik} g_{lj} \Braket{\lambda_i|\Lambda|\lambda_l}=0, 
\hspace{4mm} 
j,k=1,2, \ldots, 2N. 
\label{g_jk}
\end{align}
Thus, \mbox{$g_{jk} + g_{kj}=0$} and (\ref{g_jk}) can be 
written as a \mbox{$2N \times 2N$} matrix equation: 
\begin{equation}
GA-AG-G L G =O, 
\nonumber
\end{equation}
which is equivalent to 
\begin{equation}
AG^{-1} -G^{-1} A =L. 
\label{G_lin}
\end{equation}
Here, $A$ is a 
diagonal matrix, 
$G$ is a 
skew-symmetric matrix 
and $L$ is a symmetric matrix, 
defined as 
\begin{align}
A &:= \mathrm{diag} \left( \lambda_1, \lambda_2, \ldots, \lambda_{2N} \right), 
\nonumber \\[1mm]
G &:= \left( g_{jk} \right)_{j,k=1,2,\ldots, 2N}, 
\nonumber \\[1mm]
L &:= \left( \Braket{\lambda_l|\Lambda|\lambda_i} \right)_{l,i=1,2,\ldots, 2N}.
\nonumber 
\end{align}
By solving the linear 
equation (\ref{G_lin}) for $G^{-1}$, 
we find that 
off-diagonal entries 
of the skew-symmetric matrix $G^{-1}$ 
are 
given by 
\begin{equation}
\left( G^{-1} \right)_{jk} 
 = \frac{\Braket{\lambda_j|\Lambda|\lambda_k}}{\lambda_j-\lambda_k}, 
\hspace{4mm}
j \neq k.
\label{G_jk}
\end{equation}
In view of
(\ref{Jordan_U}) at \mbox{$\lambda=\lambda_k$} and 
(\ref{adj}) at \mbox{$\lambda=\lambda_j$}, 
we can compute 
the $x$-derivative 
of $G^{-1}$ as 
\begin{align}
\partial_x \left( G^{-1} \right)_{jk} 
&= - \left( G^{-1} G_x G^{-1} \right)_{jk}
\nonumber \\ 
&= \Braket{\lambda_j|\Lambda \, \mathrm{diag} 
	\left( \mathrm{i}, 0, \ldots, 0, -\mathrm{i} \right) 
|\lambda_k}. 
\label{G_x}
\end{align}
With the aid of 
(\ref{g_jk}) and (\ref{G_x}), 
we can prove a multifold generalization 
of Proposition~\ref{prop3.1} 
by a direct calculation. 

\begin{proposition}
\label{prop3.2}
The spectral problem \mbox{$(\ref{Jordan_U})$} is form-invariant 
under the action of the $N$-fold 
binary Darboux transformation
defined as 
\begin{equation}
\left[
\begin{array}{c}
 \widetilde{\psi}_1  \\
 \widetilde{\vt{\psi}}_2 \\
 \widetilde{\psi}_3 \\
\end{array}
\right] 
\propto
\left\{ 
I +\sum_{k=1}^{2N}  \frac{1}{\lambda-\lambda_k} \left( \sum_{j=1}^{2N} g_{jk} 
\Ket{\lambda_j} \right) \Bra{\lambda_k} \Lambda \right\}
\left[
\begin{array}{c}
 \psi_1  \\
 \vt{\psi}_2 \\
 \psi_3 \\
\end{array}
\right], 
\nonumber 
\end{equation}
up to an overall constant, 
where \mbox{$\Ket{\lambda_j}$} is a linear eigenfunction 
of the original spectral problem \mbox{$(\ref{Jordan_U})$} at 
\mbox{$\lambda=\lambda_j$} satisfying \mbox{$\Braket{\lambda_j|\Lambda|\lambda_j} =0$}
and $g_{jk}$ is the $(j,k)$ element of the inverse of 
the skew-symmetric matrix $G^{-1}$ 
determined 
by \mbox{$(\ref{G_jk})$}. 
The 
transformed 
potentials $\widetilde{\vt{q}}$ and $\widetilde{\vt{r}}$ 
are 
given 
by 
%
\begin{align}
\widetilde{\vt{q}} & = \vt{q} - \mathrm{i} 
\sum_{1 \le j < k \le 2N} g_{jk} \left( 
\Ket{\lambda_j} \Bra{\lambda_k} - \Ket{\lambda_k} \Bra{\lambda_j} 
\right)_{12}, 
\nonumber \\[2mm]
\widetilde{\vt{r}} &= \vt{r} + \mathrm{i} 
\sum_{1 \le j < k \le 2N} g_{jk} \left( 
\Ket{\lambda_j} \Bra{\lambda_k} - \Ket{\lambda_k} \Bra{\lambda_j} 
\right)_{32}, 
\nonumber 
\end{align}
where 
%
the subscripts 
${}_{12}$ and ${}_{32}$ denote 
the 
$(1,2)$ and $(3,2)$ 
sub-matrices (row vectors in this case) 
in 
the \mbox{$3 \times 3$} block matrix. 
\end{proposition}

Note that 
overall factors of 
\mbox{$\Ket{\lambda_1}, \Ket{\lambda_2}, \ldots, \Ket{\lambda_{2N}}$} 
are irrelevant to 
the 
definition of 
the $N$-fold binary Darboux transformation. 

\subsection{Multisoliton solutions}

We first notice that the spectral problem \mbox{$(\ref{Jordan_U})$} 
under the complex conjugation reduction \mbox{$\vt{r}=-\vt{q}^\ast$} 
has the following symmetry property (a kind of involution): 
if 
\[
\left[
\begin{array}{c}
 \psi_1  \\
 \vt{\psi}_2 \\
 \psi_3 \\
\end{array}
\right]
\]
is a linear eigenfunction at 
\mbox{$\lambda=\mu
$}, then 
\[
\left[
\begin{array}{c}
 \psi_3^\ast  \\
 -\vt{\psi}_2^\ast \\
 \psi_1^\ast \\
\end{array}
\right] = \Lambda 
\left[
\begin{array}{c}
 \psi_1^\ast  \\
 \vt{\psi}_2^\ast \\
 \psi_3^\ast \\
\end{array}
\right]
\]
is 
a linear eigenfunction at \mbox{$\lambda=\mu^\ast
$}. 
By applying Proposition~\ref{prop3.1} using these two linear 
eigenfunctions as $\Ket{\mu}$ and $\Ket{\nu}$, 
we obtain new 
potentials $\widetilde{\vt{q}}$ and 
$\widetilde{\vt{r}}$, 
which also satisfy the same 
relation 
\mbox{$\widetilde{\vt{r}}= -\widetilde{\vt{q}}^\ast$}. 

To obtain the bright $N$-soliton solution of the Kulish--Sklyanin model 
((\ref{nonreduceKS2}) under the reduction \mbox{$\vt{r}=-\vt{q}^\ast$}):
\begin{equation} 
\label{reducedKS2}
\mathrm{i} \vt{q}_t + \vt{q}_{xx} + 2 \sca{\vt{q}}{\vt{q}^\ast} \vt{q}
	- \sca{\vt{q}}{\vt{q}} \vt{q}^\ast = \vt{0},
\end{equation}
we start with the trivial zero solution \mbox{$\vt{q}=\vt{r}=\vt{0}$}  
and apply Proposition~\ref{prop3.2}. 
In view of the above symmetry property, 
we consider 
a 
set of $2N$ eigenvalues \mbox{$\{ \lambda_1, \lambda_2, \ldots, \lambda_{2N}\}$} 
that consist of 
$N$ complex conjugate pairs. 
The 
ordering 
of the $2N$ eigenvalues is irrelevant 
to the definition of the 
$N$-fold binary Darboux transformation, 
so 
it 
can be altered 
depending on one's preference; 
in this paper, we 
number the $2N$ eigenvalues as 
\[
\lambda_{N+j} = \lambda_{j}^\ast, \hspace{5mm} j=1,2,\ldots, N, 
\]
and choose 
a column-vector eigenfunction $\Ket{\lambda_j}$ of the linear problem 
\mbox{$(\ref{Jordan_U})$} and \mbox{$(\ref{Jordan_V})$} 
at \mbox{$\lambda=\lambda_j$} as 
\begin{subequations}
\label{eigenfunctions}
\begin{equation}
\Ket{\lambda_j} = 
\left[
\begin{array}{c}
 \mathrm{e}^{-\mathrm{i} \lambda_j x -\mathrm{i} \lambda_j^2 t} \\
 \vt{c}_j^T \\
 \frac{1}{2} \sca{\vt{c}_j}{\vt{c}_j} 
	\mathrm{e}^{\mathrm{i} \lambda_j x + \mathrm{i} \lambda_j^2 t} \\
\end{array}
\right] 
\propto 
\left[
\begin{array}{c}
 1 \\
 \vt{c}_j^T \mathrm{e}^{\mathrm{i} \lambda_j x + \mathrm{i} \lambda_j^2 t}\\
 \frac{1}{2} \sca{\vt{c}_j}{\vt{c}_j} 
	\mathrm{e}^{2\mathrm{i} \lambda_j x + 2\mathrm{i} \lambda_j^2 t} \\
\end{array}
\right], \hspace{5mm} j=1,2,\ldots, N,
\end{equation}
and 
\begin{equation}
\Ket{\lambda_{N+j}} = 
\left[
\begin{array}{c}
  \frac{1}{2} \sca{\vt{c}_j^\ast}{\vt{c}_j^\ast} 
	\mathrm{e}^{-\mathrm{i} \lambda_j^\ast x 
	- \mathrm{i} \lambda_j^{\ast \hspace{1pt} 2} t} \\
 -\vt{c}_j^\dagger \\
 \mathrm{e}^{\mathrm{i} \lambda_j^\ast x 
	+\mathrm{i} \lambda_j^{\ast \hspace{1pt} 2} t} \\
\end{array}
\right] 
\propto 
\left[
\begin{array}{c}
  \frac{1}{2} \sca{\vt{c}_j^\ast}{\vt{c}_j^\ast} 
	\mathrm{e}^{-2\mathrm{i} \lambda_j^\ast x 
	- 2\mathrm{i} \lambda_j^{\ast \hspace{1pt} 2} t} \\
 -\vt{c}_j^\dagger \mathrm{e}^{-\mathrm{i} \lambda_j^\ast x 
	- \mathrm{i} \lambda_j^{\ast \hspace{1pt} 2} t} \\
 1 \\
\end{array}
\right], \hspace{5mm} j=1,2,\ldots, N, 
\end{equation}
\end{subequations}
where $\vt{c}_j$ is a constant row vector. 
Note that 
these linear eigenfunctions 
indeed 
satisfy 
the condition \mbox{$\Braket{\lambda_j|\Lambda|\lambda_j} =0$}, $\hspace{1pt}$
\mbox{$j=1,2,\ldots, 2N$}. 

Recalling that 
overall factors of 
\mbox{$\Ket{\lambda_1}, \Ket{\lambda_2}, \ldots, \Ket{\lambda_{2N}}$} 
are irrelevant 
in 
the $N$-fold binary Darboux transformation, 
we can 
rescale 
these 
eigenfunctions as in (\ref{eigenfunctions})
and translate the skew-symmetric matrix $G^{-1}$ 
determined by \mbox{$(\ref{G_jk})$} into 
a slightly simpler skew-symmetric matrix: 
\begin{equation}
G^{-1} \to 
\left[
\begin{array}{cc}
 U & V \\
 -V^T & W \\
\end{array}
\right], \hspace{5mm} U^T = -U, 
\hspace{5mm} W^T=-W,  
\nonumber 
\end{equation}
where the 
entries of the \mbox{$N \times N$} 
matrices 
\mbox{$U:= \left( u_{jk} \right)_{j,k=1,2,\ldots, N}$}, 
\mbox{$V:= \left( v_{jk} \right)_{j,k=1,2,\ldots, N}$} and 
\mbox{$W:= \left( w_{jk} \right)_{j,k=1,2,\ldots, N}$} are defined as
%
\begin{subequations}
\label{uvw_def}
\begin{align}
 u_{jk} &:= \frac{ \frac{1}{2} \sca{\vt{c}_j}{\vt{c}_j} 
	\mathrm{e}^{2\mathrm{i} \lambda_j x + 2\mathrm{i} \lambda_j^2 t}
	+  \frac{1}{2} \sca{\vt{c}_k}{\vt{c}_k} 
	\mathrm{e}^{2\mathrm{i} \lambda_k x + 2\mathrm{i} \lambda_k^2 t} 
	- \sca{\vt{c}_j}{\vt{c}_k} 
	\mathrm{e}^{\mathrm{i} \left( \lambda_j +  \lambda_k \right) 
	x + \mathrm{i} \left( \lambda_j^2 + \lambda_k^2 \right) t}
	}{\lambda_j-\lambda_k}
\nonumber \\
& \phantom{:}= \frac{\sca{\vt{c}_j \mathrm{e}^{\mathrm{i} \lambda_j x 
	+ \mathrm{i} \lambda_j^2 t}-\vt{c}_k \mathrm{e}^{\mathrm{i} \lambda_k x 
	+ \mathrm{i} \lambda_k^2 t}}{\; \vt{c}_j \mathrm{e}^{\mathrm{i} \lambda_j x 
	+ \mathrm{i} \lambda_j^2 t}-\vt{c}_k \mathrm{e}^{\mathrm{i} \lambda_k x 
	+ \mathrm{i} \lambda_k^2 t}}
}{2\left( \lambda_j-\lambda_k \right)}, \hspace{5mm} j < 
k, 
\\
 v_{jk} &:= \frac{1 + \sca{\vt{c}_j}{\vt{c}_k^\ast} 
	\mathrm{e}^{\mathrm{i} \left( \lambda_j - \lambda_k^\ast \right) x 
	+ \mathrm{i}\left( \lambda_j^2 -\lambda_k^{\ast \hspace{1pt} 2} \right) t} 
	+ \frac{1}{4} \sca{\vt{c}_j}{\vt{c}_j} 
	\sca{\vt{c}_k^\ast}{\vt{c}_k^\ast} 
	\mathrm{e}^{2\mathrm{i} \left( \lambda_j - \lambda_k^\ast \right) x 
	+ 2\mathrm{i}\left( \lambda_j^2 -\lambda_k^{\ast \hspace{1pt} 2} \right) t} 
	}{\lambda_j-\lambda_k^\ast},
\\
 w_{jk} &:= \frac{\frac{1}{2} \sca{\vt{c}_j^\ast}{\vt{c}_j^\ast} 
	\mathrm{e}^{-2\mathrm{i} \lambda_j^\ast x 
	- 2\mathrm{i} \lambda_j^{\ast \hspace{1pt} 2} t}
	+ \frac{1}{2} \sca{\vt{c}_k^\ast}{\vt{c}_k^\ast} 
	\mathrm{e}^{-2\mathrm{i} \lambda_k^\ast x 
	- 2\mathrm{i} \lambda_k^{\ast \hspace{1pt} 2} t}
	- \sca{\vt{c}_j^\ast}{\vt{c}_k^\ast} \mathrm{e}^{-\mathrm{i} 
	\left( \lambda_j^\ast + \lambda_k^\ast \right) x 
	- \mathrm{i} \left( \lambda_j^{\ast \hspace{1pt} 2} 
	+ \lambda_k^{\ast \hspace{1pt} 2} \right) t} 
	}{\lambda_j^\ast-\lambda_k^\ast}, \hspace{5mm} j < 
k.
\nonumber
\end{align}
\end{subequations}
Note that \mbox{$u_{jj}=w_{jj}=0$} and 
$u_{jk}$ and $w_{jk}$ for \mbox{$j > k$} are 
given by $-u_{kj}$ and $-w_{kj}$, respectively. 
Moreover, 
we have \mbox{$w_{jk}=u_{jk}^\ast$} and \mbox{$v_{jk}^\ast=-v_{kj}$}, 
so \mbox{$W=U^\ast$} and \mbox{$V^\dagger=-V$}.

Now, by applying Proposition~\ref{prop3.2}, we obtain 
%
\begin{align}
\widetilde{\vt{q}} & 
=- \mathrm{i} 
\sum_{1 \le j < k \le N} 
\left[
\begin{array}{cc}
 U & V \\
 V^\ast & U^\ast \\
\end{array}
\right]^{-1}_{\;\, j k}
\left( \vt{c}_k \mathrm{e}^{\mathrm{i} \lambda_k x 
	+ \mathrm{i} \lambda_k^2 t} - \vt{c}_j \mathrm{e}^{\mathrm{i} \lambda_j x 
	+ \mathrm{i} \lambda_j^2 t} \right)
\nonumber \\ & \hphantom{=} \; \mbox{}
- \mathrm{i} 
\sum_{1 \le j, k \le N} 
\left[
\begin{array}{cc}
 U & V \\
 V^\ast & U^\ast \\
\end{array}
\right]^{-1}_{\;\, j,N+k}
 \left( -\vt{c}_k^\ast \mathrm{e}^{-\mathrm{i} \lambda_k^\ast x 
	- \mathrm{i} \lambda_k^{\ast \hspace{1pt} 2} t}
	- \frac{1}{2} \sca{\vt{c}_k^\ast}{\vt{c}_k^\ast} \vt{c}_j 
	\mathrm{e}^{ \mathrm{i} \left( \lambda_j -2 \lambda_k^\ast \right) x 
	+ \mathrm{i} \left( \lambda_j^2 - 2 \lambda_k^{\ast \hspace{1pt} 2} \right) t} 
	 \right)
\nonumber \\ & 
\hphantom{=} \; \mbox{}
- \mathrm{i} 
\sum_{1 \le j < k \le N} 
\left[
\begin{array}{cc}
 U & V \\
 V^\ast & U^\ast \\
\end{array}
\right]^{-1}_{\;\, N+j,N+k}
\left( -\frac{1}{2} \sca{\vt{c}_j^\ast}{\vt{c}_j^\ast} 
	\vt{c}_k^\ast \mathrm{e}^{-\mathrm{i} \left( 2\lambda_j^\ast 
	+ \lambda_k^\ast \right) x 
	- \mathrm{i} \left( 2 \lambda_j^{\ast \hspace{1pt} 2} 
	+ \lambda_k^{\ast \hspace{1pt} 2} \right) t} 
	 \right.
\nonumber \\
& \hspace{25mm} \left. \mbox{}
	+\frac{1}{2} \sca{\vt{c}_k^\ast}{\vt{c}_k^\ast} 
	\vt{c}_j^\ast \mathrm{e}^{-\mathrm{i} \left( 
	\lambda_j^\ast + 2 \lambda_k^\ast \right) x 
	- \mathrm{i} \left( \lambda_j^{\ast \hspace{1pt} 2} 
	+ 2 \lambda_k^{\ast \hspace{1pt} 2} \right) t} 
\right), 
\label{N-soliton1}
\end{align}
%
and \mbox{$\widetilde{\vt{r}}=-\widetilde{\vt{q}}^\ast$}, 
so the complex conjugation reduction is 
realized. 
It only remains to compute the 
entries of the inverse matrix in 
(\ref{N-soliton1}). 
%
While the inverse of a general 
square matrix is given in terms of the 
determinant and 
cofactors~\cite{Satake}, 
the inverse of a skew-symmetric matrix 
can be expressed more simply 
in terms of the Pfaffian and 
cofactors~\cite{Hirota04}. 
The Pfaffian~\cite{Satake,Hirota04} 
is 
a square root of the determinant of 
a skew-symmetric matrix of even dimension
(see https://en.wikipedia.org/wiki/Pfaffian).
For example, the inverse of 
a \mbox{$4 \times 4$} skew-symmetric matrix
is given as 
\[
\left[
\begin{array}{cccc}
 0 & d_{12} & d_{13} & d_{14} \\
 -d_{12} & 0 & d_{23} & d_{24} \\
 -d_{13} & -d_{23} & 0 & d_{34} \\
 -d_{14} & -d_{24} & -d_{34} & 0 \\
\end{array}
\right]^{-1}
= \frac{1}{d_{12}d_{34}-d_{13}d_{24}+d_{14}d_{23}}
\left[
\begin{array}{cccc}
 0 & -d_{34} & d_{24} & -d_{23} \\
 d_{34} & 0 & -d_{14} & d_{13} \\
 -d_{24} & d_{14} & 0 & -d_{12} \\
 d_{23} & -d_{13} & d_{12} & 0 \\
\end{array}
\right]. 
\]
%
Following 
the notation and definition 
in 
Hirota's book~\cite{Hirota04}, 
we write 
the Pfaffian 
of the \mbox{$2N \times 2N$} skew-symmetric matrix 
\begin{equation}
\left[
\begin{array}{cc}
 U & V \\
 V^\ast & U^\ast \\
\end{array}
\right], \hspace{5mm} U^T = -U, \hspace{5mm} V^\dagger=-V
\label{def_Pf}
\end{equation}
as 
\[
\left(1, 2, \ldots, 2N \right)
\]
and 
denote 
cofactors as 
\begin{equation}
\Gamma (j,k) = (-1)^{j+k-1} 
	\left(1, 2, \ldots, j-1, j+1, \ldots, k-1, k+1, \ldots, 2N \right), 
\hspace{5mm} 1 \le j<k \le 2N. 
\label{Gamma_def}
\end{equation}
Then, we have 
\[
\left[
\begin{array}{cc}
 U & V \\
 V^\ast & U^\ast \\
\end{array}
\right]^{-1} 
= \frac{1}{\left(1, 2, \ldots, 2N \right)}
\left[
\begin{array}{ccccc}
 0 & -\Gamma (1,2) & -\Gamma (1,3) & \cdots & -\Gamma (1,2N) \\
 \Gamma (1,2) & 0 & -\Gamma (2,3) & \cdots & -\Gamma (2,2N) \\
 \Gamma (1,3) & \Gamma (2,3) & 0 & \cdots & -\Gamma (3,2N) \\
 \vdots & \vdots & \vdots & \ddots & \vdots \\ 
 \Gamma (1,2N) & \Gamma (2,2N) & \Gamma (3,2N) & \cdots & 0 \\ 
\end{array}
\right]. 
\]
Using this formula in (\ref{N-soliton1}) and omitting the tilde 
of $\widetilde{\vt{q}}$, we
arrive at the main result of this paper. 
%
\begin{proposition}
The bright $N$-soliton solution of the self-focusing Kulish--Sklyanin model 
\mbox{$(\ref{reducedKS2})$} is given by 
\begin{align}
\vt{q} & 
= \frac{\mathrm{i}}{\left(1, 2, \ldots, 2N \right)} \left\{
\sum_{1 \le j < k \le N} \Gamma (j,k)
\left( \vt{c}_k \mathrm{e}^{\mathrm{i} \lambda_k x 
	+ \mathrm{i} \lambda_k^2 t} - \vt{c}_j \mathrm{e}^{\mathrm{i} \lambda_j x 
	+ \mathrm{i} \lambda_j^2 t} \right) \right.
\nonumber \\ & \hphantom{=} \; \mbox{}
+ \sum_{1 \le j, k \le N} 
 \Gamma (j,N+k)
 \left( -\vt{c}_k^\ast \mathrm{e}^{-\mathrm{i} \lambda_k^\ast x 
	- \mathrm{i} \lambda_k^{\ast \hspace{1pt} 2} t}
	- \frac{1}{2} \sca{\vt{c}_k^\ast}{\vt{c}_k^\ast} \vt{c}_j 
	\mathrm{e}^{ \mathrm{i} \left( \lambda_j -2 \lambda_k^\ast \right) x 
	+ \mathrm{i} \left( \lambda_j^2 - 2 \lambda_k^{\ast \hspace{1pt} 2} \right) t} 
	 \right)
\nonumber \\ & 
\hphantom{=} \; \mbox{}
+ \sum_{1 \le j < k \le N} 
\Gamma (N+j,N+k)
\left( -\frac{1}{2} \sca{\vt{c}_j^\ast}{\vt{c}_j^\ast} 
	\vt{c}_k^\ast 
	\mathrm{e}^{-\mathrm{i} \left( 2\lambda_j^\ast 
	+ \lambda_k^\ast \right) x 
	- \mathrm{i} \left( 2 \lambda_j^{\ast \hspace{1pt} 2} 
	+ \lambda_k^{\ast \hspace{1pt} 2} \right) t} 
	\right.
\nonumber \\
& \hspace{25mm} \left. \left. \mbox{}
	+\frac{1}{2} \sca{\vt{c}_k^\ast}{\vt{c}_k^\ast} 
	\vt{c}_j^\ast \mathrm{e}^{-\mathrm{i} \left( 
	\lambda_j^\ast + 2 \lambda_k^\ast \right) x 
	- \mathrm{i} \left( \lambda_j^{\ast \hspace{1pt} 2} 
	+ 2 \lambda_k^{\ast \hspace{1pt} 2} \right) t} 
\right) \right\}, 
\label{N-soliton2}
\end{align}
where the Pfaffian \mbox{$\left(1, 2, \ldots, 2N \right)$} and 
the cofactors \mbox{$\Gamma (j,k)$} are defined from 
the skew-symmetric matrix in \mbox{$(\ref{def_Pf})$} 
with the entries of $U$ and $V$ 
given by \mbox{$(\ref{uvw_def})$}. 
\end{proposition}
%

By extending 
the 
definition of the cofactors in (\ref{Gamma_def}) 
as 
\mbox{$\Gamma (k,j) = -\Gamma (j,k)$} 
and \mbox{$\Gamma (j,j) = 0$}~\cite{Hirota04}, 
we can rewrite 
(\ref{N-soliton2}) 
more concisely 
as 
\begin{align}
\vt{q} & 
= -\frac{\mathrm{i}}{\left(1, 2, \ldots, 2N \right)} \left\{
\sum_{j=1}^{N} \sum_{k=1}^{N} \left(  \Gamma (j,k) 
 + \Gamma (j,N+k)
 \frac{1}{2} \sca{\vt{c}_k^\ast}{\vt{c}_k^\ast} 
	\mathrm{e}^{-2 \mathrm{i} \lambda_k^\ast x 
	- 2 \mathrm{i} \lambda_k^{\ast \hspace{1pt} 2} t} \right)
	\vt{c}_j \mathrm{e}^{\mathrm{i} \lambda_j x 
	+ \mathrm{i} \lambda_j^2 t} 
\right.
\nonumber \\ & \hphantom{=} \, \left. \mbox{}
+ \sum_{j=1}^{N} \sum_{k=1}^{N} 
 \left( \Gamma (j,N+k) + \Gamma (N+j,N+k) \frac{1}{2} \sca{\vt{c}_j^\ast}{\vt{c}_j^\ast} 
	\mathrm{e}^{-2 \mathrm{i} \lambda_j^\ast x 
	- 2 \mathrm{i} \lambda_j^{\ast \hspace{1pt} 2} t}
	 \right) \vt{c}_k^\ast \mathrm{e}^{-\mathrm{i} \lambda_k^\ast x 
	- \mathrm{i} \lambda_k^{\ast \hspace{1pt} 2} t} \right\}
\nonumber \\[1mm]
& 
= -\frac{\mathrm{i}}{\left(1, 2, \ldots, 2N \right)} \left\{
\sum_{j=1}^{N} \left(1, 2, \ldots, j-1, \beta, j+1, \ldots, 2N \right) 
	\vt{c}_j \mathrm{e}^{\mathrm{i} \lambda_j x 
	+ \mathrm{i} \lambda_j^2 t} \right.
\nonumber \\
& 
 \hspace{15mm} \left. \mbox{}
	- \sum_{k=1}^{N} \left(1, 2, \ldots,N+k-1, \beta, N+k+1, \ldots, 2N \right) 
	\vt{c}_k^\ast \mathrm{e}^{-\mathrm{i} \lambda_k^\ast x 
	- \mathrm{i} \lambda_k^{\ast \hspace{1pt} 2} t}
	\right\},
\nonumber 
\end{align}
with \mbox{$\left( \beta, k \right)
=1$}, 
\mbox{$\, \left( \beta, N+k \right)
=\frac{1}{2} \sca{\vt{c}_k^\ast}{\vt{c}_k^\ast} \mathrm{e}^{-2 \mathrm{i} \lambda_k^\ast x 
	- 2 \mathrm{i} \lambda_k^{\ast \hspace{1pt} 2} t}$} for 
\mbox{$k=1, 2, \ldots, N$}.

By setting \mbox{$N=1$}, 
we obtain the one-soliton solution of the Kulish--Sklyanin model 
(\ref{reducedKS2}) as 
\[
\vt{q} = \frac{-\mathrm{i} \left( \lambda_1-\lambda_1^\ast \right)
	\left( \vt{c}_1^\ast \mathrm{e}^{-\mathrm{i} \lambda_1^\ast x 
	- \mathrm{i} \lambda_1^{\ast \hspace{1pt} 2} t}
	+\frac{1}{2} \sca{\vt{c}_1^\ast}{\vt{c}_1^\ast} \vt{c}_1 
	\mathrm{e}^{ \mathrm{i} \left( \lambda_1 -2 \lambda_1^\ast \right) x 
	+ \mathrm{i} \left( \lambda_1^2 - 2 \lambda_1^{\ast \hspace{1pt} 2} \right) t} 
	\right)}
	{1 + \sca{\vt{c}_1}{\vt{c}_1^\ast} 
	\mathrm{e}^{\mathrm{i} \left( \lambda_1 - \lambda_1^\ast \right) x 
	+ \mathrm{i}\left( \lambda_1^2 -\lambda_1^{\ast \hspace{1pt} 2} \right) t} 
	+ \frac{1}{4} \sca{\vt{c}_1}{\vt{c}_1} 
	\sca{\vt{c}_1^\ast}{\vt{c}_1^\ast} 
	\mathrm{e}^{2\mathrm{i} \left( \lambda_1 - \lambda_1^\ast \right) x 
	+ 2\mathrm{i}\left( \lambda_1^2 -\lambda_1^{\ast \hspace{1pt} 2} \right) t} 
	}. 
\]
The case \mbox{$\sca{\vt{c}_1}{\vt{c}_1}=0$} 
and the case \mbox{$\sca{\vt{c}_1}{\vt{c}_1} \neq 0$} 
correspond to the rank-1 one-soliton solution (\ref{rank1}) and 
the rank-2 one-soliton solution (\ref{rank2})
respectively, up to a rescaling of $\vt{q}$. 
The one- and two-soliton solutions of the Kulish--Sklyanin model (\ref{reducedKS2}) 
(up to a linear transformation mixing the components) 
have been studied in detail in~\cite{DM2010,Ieda04-1,Ieda04-2,Kanna11}. 
Note that 
the $N$-soliton solution (\ref{N-soliton2}) 
is a linear combination of 
the $2N$ 
constant vectors \mbox{$\vt{c}_1, 
\ldots, \vt{c}_N, 
\vt{c}_1^\ast, \ldots, \vt{c}_N^\ast$}, 
so it is mathematically 
redundant to consider the case where 
the number of 
the 
components of 
$\vt{q}$ 
is 
more than $2N$. 

Let us move on to the solutions of the vector mKdV equation
(\ref{vmKdV3}). 
We first notice that the spectral problem \mbox{$(\ref{Jordan_U2})$}, 
i.e., \mbox{$(\ref{Jordan_U})$} 
under the 
reduction \mbox{$\vt{r}=-\vt{q}$}, 
has the following symmetry property (a kind of involution): 
if 
\[
\left[
\begin{array}{c}
 \psi_1  \\
 \vt{\psi}_2 \\
 \psi_3 \\
\end{array}
\right]
\]
is a linear eigenfunction at 
\mbox{$\lambda=\mu
$}, then 
\[
\left[
\begin{array}{c}
 \psi_3 \\
 -\vt{\psi}_2 \\
 \psi_1 \\
\end{array}
\right] = \Lambda 
\left[
\begin{array}{c}
 \psi_1  \\
 \vt{\psi}_2 \\
 \psi_3 \\
\end{array}
\right]
\]
is 
a linear eigenfunction at \mbox{$\lambda=-\mu$}. 
By applying Proposition~\ref{prop3.1} using these two linear 
eigenfunctions as $\Ket{\mu}$ and $\Ket{\nu}$, 
we obtain new 
potentials $\widetilde{\vt{q}}$ and 
$\widetilde{\vt{r}}$, 
which also satisfy the same 
relation 
\mbox{$\widetilde{\vt{r}}= -\widetilde{\vt{q}}$}. 

To obtain the 
multisoliton (or multi-breather) solutions of 
%
the vector mKdV equation
(\ref{vmKdV3}), 
we start with the trivial zero solution \mbox{$\vt{q}
=\vt{r}
=\vt{0}$} in the spectral problem \mbox{$(\ref{Jordan_U})$} 
and  apply Proposition~\ref{prop3.2}. 
In view of the above symmetry property, 
we consider 
the case where the $2N$ eigenvalues 
\mbox{$\{ \lambda_1, \lambda_2, \ldots, \lambda_{2N}\}$} 
occur in plus-minus pairs. 
The 
ordering 
of the $2N$ eigenvalues is irrelevant 
to the definition of the 
$N$-fold binary Darboux transformation
and 
can be altered; 
in this paper, we 
number the $2N$ eigenvalues as 
\[
\lambda_{N+j} = -\lambda_{j}, \hspace{5mm} j=1,2,\ldots, N, 
\]
and choose 
a column-vector eigenfunction $\Ket{\lambda_j}$ of the linear problem 
\mbox{$(\ref{Jordan_U2})$} and \mbox{$(\ref{Jordan_V2})$} 
at \mbox{$\lambda=\lambda_j$} as 
\begin{subequations}
\label{eigenfunctions2}
\begin{equation}
\Ket{\lambda_j} = 
\left[
\begin{array}{c}
 \mathrm{e}^{-\mathrm{i} \lambda_j x -\mathrm{i} \lambda_j^3 y} \\
 \vt{c}_j^T \\
 \frac{1}{2} \sca{\vt{c}_j}{\vt{c}_j} 
	\mathrm{e}^{\mathrm{i} \lambda_j x + \mathrm{i} \lambda_j^3 y} \\
\end{array}
\right] 
\propto 
\left[
\begin{array}{c}
 1 \\
 \vt{c}_j^T \mathrm{e}^{\mathrm{i} \lambda_j x + \mathrm{i} \lambda_j^3 y}\\
 \frac{1}{2} \sca{\vt{c}_j}{\vt{c}_j} 
	\mathrm{e}^{2\mathrm{i} \lambda_j x + 2\mathrm{i} \lambda_j^3 y} \\
\end{array}
\right], \hspace{5mm} j=1,2,\ldots, N,
\label{eigenfunctions2-1}
\end{equation}
and 
\begin{equation}
\Ket{\lambda_{N+j}} = \Lambda \Ket{\lambda_j} 
\propto 
\left[
\begin{array}{c}
  \frac{1}{2} \sca{\vt{c}_j}{\vt{c}_j} 
	\mathrm{e}^{2\mathrm{i} \lambda_j x 
	+ 2\mathrm{i} \lambda_j^3 y} \\
 -\vt{c}_j^T \mathrm{e}^{\mathrm{i} \lambda_j x 
	+ \mathrm{i} \lambda_j^3 y} \\
 1 \\
\end{array}
\right], \hspace{5mm} j=1,2,\ldots, N, 
\end{equation}
\end{subequations}
where $\vt{c}_j$ is a constant row vector. 
Note that 
these linear eigenfunctions 
indeed 
satisfy 
the condition \mbox{$\Braket{\lambda_j|\Lambda|\lambda_j} =0$}, $\hspace{1pt}$
\mbox{$j=1,2,\ldots, 2N$}. 

Recalling that 
overall factors of 
\mbox{$\Ket{\lambda_1}, \Ket{\lambda_2}, \ldots, \Ket{\lambda_{2N}}$} 
are irrelevant 
in 
the $N$-fold binary Darboux transformation, 
we can 
rescale 
these 
eigenfunctions as in (\ref{eigenfunctions2})
and translate the skew-symmetric matrix $G^{-1}$ 
determined by \mbox{$(\ref{G_jk})$} into 
a slightly simpler skew-symmetric matrix: 
\begin{equation}
G^{-1} \to 
\left[
\begin{array}{cc}
 U & V \\
 -V^T & -U \\
\end{array}
\right], \hspace{5mm} U^T = -U, 
\nonumber 
\end{equation}
where the 
entries 
of the \mbox{$N \times N$} 
matrices 
\mbox{$U:= \left( u_{jk} \right)_{j,k=1,2,\ldots, N}$} 
and \mbox{$V:= \left( v_{jk} \right)_{j,k=1,2,\ldots, N}$} 
are defined as
%
\begin{subequations}
\label{uvw_def2}
\begin{align}
 u_{jk} &:= \frac{ \frac{1}{2} \sca{\vt{c}_j}{\vt{c}_j} 
	\mathrm{e}^{2\mathrm{i} \lambda_j x + 2\mathrm{i} \lambda_j^3 y}
	+  \frac{1}{2} \sca{\vt{c}_k}{\vt{c}_k} 
	\mathrm{e}^{2\mathrm{i} \lambda_k x + 2\mathrm{i} \lambda_k^3 y} 
	- \sca{\vt{c}_j}{\vt{c}_k} 
	\mathrm{e}^{\mathrm{i} \left( \lambda_j +  \lambda_k \right) 
	x + \mathrm{i} \left( \lambda_j^3 + \lambda_k^3 \right) y}
	}{\lambda_j-\lambda_k}
\nonumber \\
& \phantom{:}= \frac{\sca{\vt{c}_j \mathrm{e}^{\mathrm{i} \lambda_j x 
	+ \mathrm{i} \lambda_j^3 y}-\vt{c}_k \mathrm{e}^{\mathrm{i} \lambda_k x 
	+ \mathrm{i} \lambda_k^3 y}}{\; \vt{c}_j \mathrm{e}^{\mathrm{i} \lambda_j x 
	+ \mathrm{i} \lambda_j^3 y}-\vt{c}_k \mathrm{e}^{\mathrm{i} \lambda_k x 
	+ \mathrm{i} \lambda_k^3 y}}
}{2\left( \lambda_j-\lambda_k \right)}, \hspace{5mm} j < 
k, 
\\
 v_{jk} &:= \frac{1 + \sca{\vt{c}_j}{\vt{c}_k} 
	\mathrm{e}^{\mathrm{i} \left( \lambda_j + \lambda_k \right) x 
	+ \mathrm{i}\left( \lambda_j^3 +\lambda_k^3 \right) y} 
	+ \frac{1}{4} \sca{\vt{c}_j}{\vt{c}_j} 
	\sca{\vt{c}_k}{\vt{c}_k} 
	\mathrm{e}^{2\mathrm{i} \left( \lambda_j + \lambda_k \right) x 
	+ 2\mathrm{i}\left( \lambda_j^3 +\lambda_k^3 \right) y} 
	}{\lambda_j+\lambda_k}.
\end{align}
\end{subequations}
Note that \mbox{$v_{jk}=v_{kj}$}, 
so 
\mbox{$V^T=V$}.

Now, by applying Proposition~\ref{prop3.2}, we obtain 
%
\begin{align}
\widetilde{\vt{q}} & 
=- \mathrm{i} 
\sum_{1 \le j < k \le N} 
\left[
\begin{array}{cc}
 U & V \\
 -V & -U \\
\end{array}
\right]^{-1}_{\;\, j k}
\left( \vt{c}_k \mathrm{e}^{\mathrm{i} \lambda_k x 
	+ \mathrm{i} \lambda_k^3 y} - \vt{c}_j \mathrm{e}^{\mathrm{i} \lambda_j x 
	+ \mathrm{i} \lambda_j^3 y} \right)
\nonumber \\ & \hphantom{=} \; \mbox{}
- \mathrm{i} 
\sum_{1 \le j, k \le N} 
\left[
\begin{array}{cc}
 U & V \\
 -V & -U \\
\end{array}
\right]^{-1}_{\;\, j,N+k}
 \left( -\vt{c}_k \mathrm{e}^{\mathrm{i} \lambda_k x 
	+ \mathrm{i} \lambda_k^3 y}
	- \frac{1}{2} \sca{\vt{c}_k}{\vt{c}_k} \vt{c}_j 
	\mathrm{e}^{ \mathrm{i} \left( \lambda_j +2 \lambda_k \right) x 
	+ \mathrm{i} \left( \lambda_j^3 + 2 \lambda_k^3 \right) y} 
	 \right)
\nonumber \\ & 
\hphantom{=} \; \mbox{}
- \mathrm{i} 
\sum_{1 \le j < k \le N} 
\left[
\begin{array}{cc}
 U & V \\
 -V & -U \\
\end{array}
\right]^{-1}_{\;\, N+j,N+k}
\left( -\frac{1}{2} \sca{\vt{c}_j}{\vt{c}_j} 
	\vt{c}_k \mathrm{e}^{\mathrm{i} \left( 2\lambda_j 
	+ \lambda_k \right) x 
	+ \mathrm{i} \left( 2 \lambda_j^3
	+ \lambda_k^3 \right) y} 
	 \right.
\nonumber \\
& \hspace{25mm} \left. \mbox{}
	+\frac{1}{2} \sca{\vt{c}_k}{\vt{c}_k} 
	\vt{c}_j \mathrm{e}^{\mathrm{i} \left( 
	\lambda_j + 2 \lambda_k \right) x 
	+ \mathrm{i} \left( \lambda_j^3
	+ 2 \lambda_k^3 \right) y} 
\right), 
\label{N-soliton3}
\end{align}
%
and \mbox{$\widetilde{\vt{r}}=-\widetilde{\vt{q}}$}, 
so the required 
reduction is 
indeed 
realized. 
Because 
\[
\left[
\begin{array}{cc}
 O & I \\
 I & O \\
\end{array}
\right]
\left[
\begin{array}{cc}
 U & V \\
 -V & -U \\
\end{array}
\right]^{-1} 
+
\left[
\begin{array}{cc}
 U & V \\
 -V & -U \\
\end{array}
\right]^{-1}
\left[
\begin{array}{cc}
 O & I \\
 I & O \\
\end{array}
\right]
 =O, 
\]
the inverse matrix should take the form:
\[
\left[
\begin{array}{cc}
 U & V \\
 -V & -U \\
\end{array}
\right]^{-1}
= 
\left[
\begin{array}{cc}
 X & Y \\
 -Y & -X \\
\end{array}
\right], \hspace{5mm} X^T=-X,  \hspace{5mm} Y^T=Y, 
\]
which is skew-symmetric. 
Thus, we can rewrite (\ref{N-soliton3}) as 
\begin{align}
\vt{q} & 
= - \mathrm{i} 
\sum_{1 \le j < k \le N} 
\left[
\begin{array}{cc}
 U & V \\
 -V & -U \\
\end{array}
\right]^{-1}_{\;\, j k}
\left( - \vt{c}_j \mathrm{e}^{\mathrm{i} \lambda_j x 
	+ \mathrm{i} \lambda_j^3 y} +\frac{1}{2} \sca{\vt{c}_j}{\vt{c}_j} 
	\vt{c}_k \mathrm{e}^{\mathrm{i} \left( 2\lambda_j 
	+ \lambda_k \right) x 
	+ \mathrm{i} \left( 2 \lambda_j^3
	+ \lambda_k^3 \right) y} 
	 \right.
\nonumber \\
& \hspace{25mm} \left. \mbox{}
	+\vt{c}_k \mathrm{e}^{\mathrm{i} \lambda_k x 
	+ \mathrm{i} \lambda_k^3 y}-\frac{1}{2} \sca{\vt{c}_k}{\vt{c}_k} 
	\vt{c}_j \mathrm{e}^{\mathrm{i} \left( 
	\lambda_j + 2 \lambda_k \right) x 
	+ \mathrm{i} \left( \lambda_j^3
	+ 2 \lambda_k^3 \right) y} 
\right)
\nonumber \\ & \hphantom{=} \; \mbox{}
- \mathrm{i} 
\sum_{1 \le j < k \le N} 
\left[
\begin{array}{cc}
 U & V \\
 -V & -U \\
\end{array}
\right]^{-1}_{\;\, j,N+k}
 \left( -\vt{c}_j \mathrm{e}^{\mathrm{i} \lambda_j x 
	+ \mathrm{i} \lambda_j^3 y}
	- \frac{1}{2} \sca{\vt{c}_j}{\vt{c}_j} \vt{c}_k 
	\mathrm{e}^{ \mathrm{i} \left( 2 \lambda_j + \lambda_k \right) x 
	+ \mathrm{i} \left( 2 \lambda_j^3 + \lambda_k^3 \right) y}  
 \right.
\nonumber \\
& \hspace{25mm} \left. \mbox{}
-\vt{c}_k \mathrm{e}^{\mathrm{i} \lambda_k x 
	+ \mathrm{i} \lambda_k^3 y}
	- \frac{1}{2} \sca{\vt{c}_k}{\vt{c}_k} \vt{c}_j 
	\mathrm{e}^{ \mathrm{i} \left( \lambda_j +2 \lambda_k \right) x 
	+ \mathrm{i} \left( \lambda_j^3 + 2 \lambda_k^3 \right) y} 
\right)
\nonumber \\ & 
\hphantom{=} \; \mbox{}
- \mathrm{i} 
\sum_{j=1}^{N} 
\left[
\begin{array}{cc}
 U & V \\
 -V & -U \\
\end{array}
\right]^{-1}_{\;\, j,N+j}
 \left( -\vt{c}_j \mathrm{e}^{\mathrm{i} \lambda_j x 
	+ \mathrm{i} \lambda_j^3 y}
	- \frac{1}{2} \sca{\vt{c}_j}{\vt{c}_j} \vt{c}_j 
	\mathrm{e}^{ 3 \mathrm{i} \lambda_j x + 3 \mathrm{i} \lambda_j^3 y} 
	 \right),
\label{N-soliton4}
\end{align}
where the tilde of $\widetilde{\vt{q}}$ is omitted. 
This is a 
fairly 
general 
complex-valued 
solution of 
the vector mKdV equation
(\ref{vmKdV3}). 
To turn 
it into 
real-valued solutions, 
we first set 
\begin{equation}
\lambda_j =  \mathrm{i} \eta_{j}, 
\hspace{5mm} j=1,2,\ldots, N, 
\label{eta_def}
\end{equation}
and rewrite the solution (\ref{N-soliton4}) as follows. 

\begin{proposition}
\label{prop3.4}
An $N$-soliton solution of the vector mKdV equation 
\mbox{$(\ref{vmKdV3})$} is given by 
\begin{align}
\vt{q} & 
= \frac{1}{\left(1, 2, \ldots, 2N \right)} \left\{ 
\sum_{1 \le j < k \le N} \Gamma (j,k) 
\left( \vt{c}_j \mathrm{e}^{-\eta_j x 
	+ \eta_j^3 y} -\frac{1}{2} \sca{\vt{c}_j}{\vt{c}_j} 
	\vt{c}_k \mathrm{e}^{-\left( 2\eta_j 
	+ \eta_k \right) x 
	+ \left( 2 \eta_j^3 + \eta_k^3 \right) y} 
	 \right. \right.
\nonumber \\
& \hspace{25mm} \left. \mbox{}
	-\vt{c}_k \mathrm{e}^{-\eta_k x 
	+ \eta_k^3 y}+\frac{1}{2} \sca{\vt{c}_k}{\vt{c}_k} 
	\vt{c}_j \mathrm{e}^{ -\left( 
	\eta_j + 2 \eta_k \right) x 
	+ \left( \eta_j^3
	+ 2 \eta_k^3 \right) y} 
\right)
\nonumber \\ & \hphantom{=} \; \mbox{}
+ \sum_{1 \le j < k \le N} \Gamma (j,N+k)
 \left( \vt{c}_j \mathrm{e}^{-\eta_j x + \eta_j^3 y}
	+ \frac{1}{2} \sca{\vt{c}_j}{\vt{c}_j} \vt{c}_k 
	\mathrm{e}^{ -\left( 2 \eta_j + \eta_k \right) x 
	+ \left( 2 \eta_j^3 + \eta_k^3 \right) y}
 \right.
\nonumber \\
& \hspace{25mm} \left. \mbox{}
 + \vt{c}_k \mathrm{e}^{-\eta_k x + \eta_k^3 y}
	+ \frac{1}{2} \sca{\vt{c}_k}{\vt{c}_k} \vt{c}_j 
	\mathrm{e}^{ - \left( \eta_j +2 \eta_k \right) x 
	+ \left( \eta_j^3 + 2 \eta_k^3 \right) y} \right)
\nonumber \\ & 
\hphantom{=} \; \left. \mbox{}
+ \sum_{j=1}^{N} \Gamma (j,N+j)
 \left( \vt{c}_j \mathrm{e}^{-\eta_j x 
	+ \eta_j^3 y}
	+ \frac{1}{2} \sca{\vt{c}_j}{\vt{c}_j} \vt{c}_j 
	\mathrm{e}^{ -3 \eta_j x + 3 \eta_j^3 y} 
	 \right) \right\},
\label{N-soliton5}
\end{align}
where 
\mbox{$\left(1, 2, \ldots, 2N \right)$} 
is 
the Pfaffian (a square root of the determinant) of 
the skew-symmetric matrix with the entries 
\begin{align}
 (j,k) &= - \left( N+j,N+k \right) 
\nonumber \\[1mm]
 &= \frac{\sca{\vt{c}_j \mathrm{e}^{- \eta_j x 
	+ \eta_j^3 y}-\vt{c}_k \mathrm{e}^{-\eta_k x 
	+ \eta_k^3 y}}{\; \vt{c}_j \mathrm{e}^{- \eta_j x 
	+ \eta_j^3 y}-\vt{c}_k \mathrm{e}^{-\eta_k x 
	+ \eta_k^3 y}}
}{2\left( \eta_j-\eta_k \right)}, 
\nonumber \\
& \hspace{91mm}
1 \le j < k \le N, 
\nonumber \\
  (j,N+k) &= \frac{1 + \sca{\vt{c}_j}{\vt{c}_k} 
	\mathrm{e}^{- \left( \eta_j + \eta_k \right) x 
	+ \left( \eta_j^3 +\eta_k^3 \right) y} 
	+ \frac{1}{4} \sca{\vt{c}_j}{\vt{c}_j} 
	\sca{\vt{c}_k}{\vt{c}_k} 
	\mathrm{e}^{-2 \left( \eta_j + \eta_k \right) x 
	+ 2 \left( \eta_j^3 +\eta_k^3 \right) y} 
	}{\eta_j+\eta_k}, 
\nonumber \\
& \hspace{91mm}
1 \le j, k \le N,
\nonumber 
\end{align}
and the cofactors \mbox{$\Gamma (j,k)$} 
for \mbox{$1 \le j<k \le 2N$} 
are defined as in \mbox{$(\ref{Gamma_def})$}.  
\end{proposition}

Note that 
(\ref{N-soliton5}) 
is of the form:\ 
\[
\vt{q} = \frac{\sum_{l=1}^{N} G_l \vt{c}_l \mathrm{e}^{-\eta_l x 
	+ \eta_l^3 y}}{F}, 
\]
where $F$ and \mbox{$G_1, \ldots, G_N$} 
are polynomials in \mbox{$
\sca{\vt{c}_j \mathrm{e}^{-\eta_j x + \eta_j^3 y}}
{\vt{c}_k \mathrm{e}^{-\eta_k x + \eta_k^3 y}} 
$} for \mbox{$1 \le j \le k \le N$}; 
this 
provides 
a real-valued $N$-soliton solution 
if 
\mbox{$\eta_1, \ldots, \eta_N 
$} are positive and 
\mbox{$ \vt{c}_1, \ldots, \vt{c}_N 
$}
are 
real. By setting \mbox{$N=1$}, we obtain 
\begin{equation}
\vt{q} = 2\eta_1 \frac{ \vt{c}_1  \mathrm{e}^{-\eta_1 x 
	+ \eta_1^3 y}}
	{1 + \frac{1}{2}\sca{\vt{c}_1}{\vt{c}_1} 
	\mathrm{e}^{- 2 \eta_1 x + 2 \eta_1^3 y}} , 
\label{scalar_soliton}
\end{equation}
which is the straightforward vector analog of the 
one-soliton solution of the scalar mKdV equation, 
i.e., the scalar 
mKdV soliton with a 
coefficient unit vector 
\mbox{$\vt{c}_1/ \sqrt{\sca{\vt{c}_1}{\vt{c}_1}}$\hspace{1pt}}. 
Incidentally, the scalar mKdV equation was 
first 
solved by 
R.~Hirota (J.\ Phys.\ Soc.\ Jpn.$\hspace{1pt}$), 
M.~Wadati (J.\ Phys.\ Soc.\ Jpn.$\hspace{1pt}$) and 
S.~Tanaka (Publ.\ RIMS $\&$ Proc.\ Japan Acad.$\hspace{1pt}$) 
almost 
independently 
in 1972. 
The solution 
(\ref{N-soliton5}) in the case of real soliton parameters 
provides 
a nontrivial vector generalization 
of the $N$-soliton solution of the scalar mKdV equation 
involving $N$ polarization vectors: 
\[
\frac{\vt{c}_1}{\sqrt{\sca{\vt{c}_1}{\vt{c}_1}}}, \;
\frac{\vt{c}_2}{\sqrt{\sca{\vt{c}_2}{\vt{c}_2}}}, \;
\ldots, \; 
\frac{\vt{c}_N}{\sqrt{\sca{\vt{c}_N}{\vt{c}_N}}} \hspace{1pt}.
\]

We can consider a 
generalization of 
the vector mKdV equation
(\ref{vmKdV3}) as considered 
by Iwao and Hirota~\cite{Iwao1997}: 
\begin{equation}
\label{vmKdV4}
\vt{q}_y + \vt{q}_{xxx} +3 \sca{\vt{q}B}{\vt{q}} \vt{q}_x = \vt{0}. 
\end{equation}
Here, 
\mbox{$B
= \left( b_{jk} \right)$} is a constant square matrix, 
which 
can be 
assumed to be symmetric 
(\mbox{$b_{jk}=b_{kj}$}) 
without loss of generality. 
Then, an $N$-soliton solution of 
this generalized 
vector mKdV equation 
(\ref{vmKdV4}) is 
given 
by 
(\ref{N-soliton5})
with 
the involved 
scalar products 
generalized 
as 
\mbox{$\sca{\vt{c}_j}{\vt{c}_k} \to \sca{\vt{c}_j B}{\vt{c}_k} $}, 
\mbox{$1 \le j, k \le N$}. 
This formula 
generalizes 
the 
multisoliton formula 
proposed 
by Iwao and Hirota~\cite{Iwao1997}  
using the Hirota 
bilinear method~\cite{Hirota04} 
(also see 
the relevant results in~\cite{Matsuno11,Feng15,Maruno15}) 
and appears to be more efficient. 
In fact, 
(\ref{N-soliton5}) 
with positive \mbox{$\eta_1, \ldots, \eta_N 
$} and real 
\mbox{$\vt{c}_1, \ldots, \vt{c}_N
$} is 
only 
a special $N$-soliton solution of the vector mKdV equation 
\mbox{$(\ref{vmKdV3})$}, which does not 
exhibit any 
oscillating behavior in each component of the vector variable $\vt{q}$.  
In particular, it cannot reproduce the
one-soliton solution 
of the complex mKdV equation~\cite{
AKNS74,Hirota73JMP,Ab78}, 
involving a complex carrier wave, 
which is equivalent 
to 
the two-component vector mKdV equation 
(\mbox{$(\ref{vmKdV3})$} with a real two-component vector $\vt{q}$), 
up to a linear transformation. 

%

To obtain the general real-valued 
multisoliton solutions of the vector mKdV equation 
(\ref{vmKdV3}), we 
require that 
\mbox{$\{ \eta_1, \eta_2, \ldots, \eta_N \}$} in (\ref{eta_def}) and (\ref{N-soliton5}), 
as well as the corresponding linear eigenfunctions in (\ref{eigenfunctions2-1}), 
are 
either real or 
occur in complex conjugate pairs~\cite{JPW01,TsuJPSJ98}. 
That is, 
up to a re-ordering, 
we assume 
\vspace{1mm}

(a) \hspace{1pt} 
\mbox{$\eta_{M+j} = \eta_{j}^\ast \;(\mathrm{Re} \, \eta_j>0), 
\; \, \vt{c}_{M+j}= \vt{c}_j^\ast, \;\, j=1, 2, \ldots, M$},
\vspace{1mm}

(b) \hspace{1pt} \mbox{$\eta_j >0, 
\;\, \vt{c}_j^\ast = \vt{c}_j, \;\, j=2M+1, 
	\ldots, 2 M+L \hspace{1pt}(=N)$}. 
\vspace{1mm}
\\
Thus, 
the original $2N$ eigenvalues 
\mbox{$\{ \lambda_1, \lambda_2, \ldots, \lambda_{2N}\}$} 
occur in (a) plus-minus and complex-conjugate quartets 
or (b) plus-minus pairs. 
Under these 
conditions, we can 
easily show 
from Proposition~\ref{prop3.2} that 
the new solution 
generated 
by the $N$-fold binary Darboux transformation 
is indeed real-valued.
%

Proposition~\ref{prop3.4} 
with the soliton parameters satisfying 
(a) and (b) 
generates a 
mixture of multisoliton and multi-breather 
solutions. To exclude breather solutions, we need only impose 
the following additional 
conditions for (a)~\cite{JPW01,TsuJPSJ98}: 
\vspace{1mm}

(a') \hspace{1pt} 
\mbox{$\sca{\vt{c}_j}{\vt{c}_j}=0, \;\, j=1, 2, \ldots, M$}. 
\vspace{1mm}
\\
Then, 
in the simplest nontrivial case of 
\mbox{$M=1$}, \mbox{$L=0$} and \mbox{$N=2$}, 
(\ref{N-soliton5}) 
gives the general one-soliton solution of the vector mKdV 
equation (\ref{vmKdV3})~\cite{TsuJMP10}: 
\begin{equation}
\vt{q} = 2 \left( \eta_1 +\eta_2 \right) 
 \frac{ \vt{c}_1 \eta_1 \mathrm{e}^{-\eta_1 x 
	+ \eta_1^3 y} -  \vt{c}_2 \eta_2 
	\mathrm{e}^{-\eta_2 x + \eta_2^3 y}}
	{\eta_1 -\eta_2 - \frac{4\eta_1 \eta_2}{\eta_1 - \eta_2}\sca{\vt{c}_1}{\vt{c}_2} 
	\mathrm{e}^{- \left( \eta_1 + \eta_2 \right) x 
	+ \left( \eta_1^3 +\eta_2^3 \right) y}
	}. 
\label{vector_soliton}
\end{equation}
With \mbox{$\eta_2 = \eta_1^\ast$}
and \mbox{$\vt{c}_2= \vt{c}_1^\ast$}, 
this is indeed a real solution and 
provides 
the 
vector analog of the complex mKdV soliton~\cite{
AKNS74,Hirota73JMP,Ab78}. 
By redefining 
the constant vector as \mbox{$2\eta_1 \vt{c}_1 =: 
\left( \eta_1-\eta_1^\ast \right) \vt{d}_1
$}, 
the one-soliton solution (\ref{vector_soliton}) can be rewritten 
in a more concise form:
\begin{equation}
\vt{q} = \left( \eta_1 +\eta_1^\ast \right) 
 \frac{ \vt{d}_1  \mathrm{e}^{-\eta_1 x 
	+ \eta_1^3 y} + \vt{d}_1^\ast 
	\mathrm{e}^{-\eta_1^\ast x + \eta_1^{\ast \hspace{1pt} 3} y}}
	{1 + \sca{\vt{d}_1}{\vt{d}_1^\ast} 
	\mathrm{e}^{- \left( \eta_1 + \eta_1^\ast \right) x 
	+ \left( \eta_1^3 +\eta_1^{\ast \hspace{1pt} 3} \right) y}
	}, \hspace{5mm} \sca{\vt{d}_1}{\vt{d}_1}=0.
\label{vector_soliton2}
\end{equation}
Thus, in 
the limit 
\mbox{$\eta_1 \to \eta_1^\ast$}, 
i.e., \mbox{$\mathrm{Im} \, \eta_1 \to 0$}, 
this solution 
reduces to (\ref{scalar_soliton}). 

For general values of $M$ and $L$, (\ref{N-soliton5}) 
with 
the above
conditions (a), (a') and (b) 
provides 
the \mbox{$(M+L)$}-soliton solution, 
a nonlinear superposition of 
$M$ vector solitons of the 
oscillating type (\ref{vector_soliton2}) 
and
$L$ vector solitons of the 
nonoscillating
type (\ref{scalar_soliton}). 
Because the nonoscillating-type 
soliton  
can be obtained 
from 
the oscillating-type 
soliton through the 
limiting procedure, 
we can 
somewhat loosely 
consider that 
the 
general 
$M$-soliton 
solution of the vector mKdV 
equation (\ref{vmKdV3}) is 
obtained by setting \mbox{$N=2M$} and \mbox{$L=0$} 
and assuming the conditions (a) and (a'). 
Thus, Proposition~\ref{prop3.4} can be restated 
as follows. 

\begin{proposition}
\label{prop3.5}
The general real-valued 
$M$-soliton solution of the vector mKdV equation 
\mbox{$(\ref{vmKdV3})$} is given by 
\begin{align}
\vt{q} & 
= \frac{1}{\left(1, 2, \ldots, 4M \right)} \left\{ 
\sum_{1 \le j < k \le 2M} \Gamma (j,k) 
\left( \vt{c}_j \mathrm{e}^{-\eta_j x 
	+ \eta_j^3 y} 
	-\vt{c}_k \mathrm{e}^{-\eta_k x 
	+ \eta_k^3 y}
\right) \right.
\nonumber \\ & \hphantom{=} \; \mbox{}
+ \sum_{1 \le j < k \le 2M} \Gamma (j,2M+k)
 \left( \vt{c}_j \mathrm{e}^{-\eta_j x + \eta_j^3 y}
 + \vt{c}_k \mathrm{e}^{-\eta_k x + \eta_k^3 y} \right)
\nonumber \\ & 
\hphantom{=} 
\left. \mbox{}
+ \sum_{j=1}^{2M} \Gamma (j,2M+j) \hspace{1pt} \vt{c}_j \mathrm{e}^{-\eta_j x 
	+ \eta_j^3 y} \right\},
\label{N-soliton6}
\end{align}
where 
\mbox{$\left(1, 2, \ldots, 4M \right)$} 
is 
the Pfaffian (a square root of the determinant) of 
the skew-symmetric matrix with the entries 
\begin{align}
 (j,k) &= - \left( 2M+j,2M+k \right) 
\nonumber \\
 &= -\frac{\sca{\vt{c}_j}{\vt{c}_k} 
	\mathrm{e}^{- \left( \eta_j + \eta_k \right) x 
	+ \left( \eta_j^3 +\eta_k^3 \right) y}
}{\eta_j-\eta_k}, 
\hspace{5mm} 1 \le j < k \le 2M, 
\nonumber \\[3mm]
  \left( j,2M+k \right) &= \left( k,2M+j \right)
\nonumber \\
&= \frac{1 + \sca{\vt{c}_j}{\vt{c}_k} 
	\mathrm{e}^{- \left( \eta_j + \eta_k \right) x 
	+ \left( \eta_j^3 +\eta_k^3 \right) y}}{\eta_j+\eta_k}, 
\hspace{5mm} 1 \le j < k \le 2M,
\nonumber \\[3mm]
  (j,2M+j) &= \frac{1}{2\eta_j}, \hspace{5mm} 1 \le j \le 2M,
\nonumber 
\end{align}
and the cofactors \mbox{$\Gamma (j,k)$} 
for \mbox{$1 \le j<k \le 4M$} 
are defined as in \mbox{$(\ref{Gamma_def})$}. 
Here, 
\mbox{$\eta_{M+j} = \eta_{j}^\ast \;(\mathrm{Re} \, \eta_j>0), 
\; \, \vt{c}_{M+j}= \vt{c}_j^\ast 
$} 
and \mbox{$\sca{\vt{c}_j}{\vt{c}_j}=0\,$}
for \mbox{$j=1, 2, \ldots, M$}. 
\end{proposition}

By extending 
the 
definition of the cofactors in (\ref{Gamma_def}) 
as 
\mbox{$\Gamma (k,j) = -\Gamma (j,k)$} 
and \mbox{$\Gamma (j,j) = 0$}~\cite{Hirota04} 
and 
noting the relation \mbox{$\Gamma (j,2M+k)=\Gamma (k,2M+j)$}, 
we can rewrite 
(\ref{N-soliton6}) 
in a more compact form as 
\begin{align}
\vt{q} & 
= \frac{1}{\left(1, 2, \ldots, 4M \right)} \sum_{j=1}^{2M} 
\left( \sum_{k=1}^{4M} \Gamma (j,k) \right)  \vt{c}_j \mathrm{e}^{-\eta_j x 
	+ \eta_j^3 y}
\nonumber \\[2mm]
& =\frac{
	\sum_{j=1}^{2M} \left(1, 2,\ldots, j-1, \beta, j+1, \ldots, 4M \right)
	\vt{c}_j \mathrm{e}^{-\eta_j x 
	+ \eta_j^3 y}}
{\left(1, 2, \ldots, 4M \right)}, 
\nonumber 
\end{align}
with \mbox{$\left( \beta, k \right)
=1, \; k=1, 2, \ldots, 4M$}. 
%
%
For the generalized 
vector mKdV equation 
(\ref{vmKdV4}) with a real symmetric and positive definite matrix $B$, 
the above formula with 
\mbox{$\sca{\vt{c}_j}{\vt{c}_k} \to \sca{\vt{c}_j B}{\vt{c}_k} $}
and \mbox{$\sca{\vt{c}_j}{\vt{c}_j}=0 \to \sca{\vt{c}_j B}{\vt{c}_j}=0 $} 
provides the general real-valued 
$M$-soliton solution. 

If we 
generalize 
the time dependence 
as 
\mbox{$\vt{c}_j \mathrm{e}^{-\eta_j x + \eta_j^3 y} \to 
\vt{c}_j \mathrm{e}^{-\eta_j x + \eta_j^3 y + \eta_j^{-1}z}$}, 
%
Propositions~\ref{prop3.5} provides 
the general real-valued 
$M$-soliton solution of the vector sine-Gordon equation~\cite{PR79,EP79} 
(up to a sign ambiguity of the square root) 
with 
the independent variables $x$ and $z$. 
Then, 
in 
the special case 
\mbox{$\vt{c}_j = \left( \vt{a}_j, \mathrm{i} \vt{a}_j \right)$} or 
\[
\vt{c}_j = \left( c_j^{(1)}, \, \mathrm{i} c_j^{(1)}, \, c_j^{(2)}, \, \mathrm{i} c_j^{(2)},\, 
\ldots \right), 
\]
the condition \mbox{$\sca{\vt{c}_j}{\vt{c}_j}=0\,$} 
is automatically satisfied 
and 
our 
$M$-soliton formula 
apparently 
reduces to 
the formula 
proposed by Feng~\cite{Feng15}. 
He 
investigated the asymptotic behavior of the 
two-soliton solution in this special case and 
showed 
that 
the two-soliton collision in the vector sine-Gordon 
equation 
is highly nontrivial, 
reflecting 
the 
internal degrees of freedom 
of the solitons. 
This 
apparently disagrees with the conclusion of~\cite{JPW01} 
that the soliton interactions in the 
vector sine-Gordon equation
are 
exactly 
the same as in
the scalar sine-Gordon equation. 
Propositions~\ref{prop3.5} could be 
used to resolve the discrepancy. 

%


%

\section{Concluding remarks}

The 
vector NLS equation
known as the Kulish--Sklyanin model~\cite{KuSk81} 
admits two different Lax representations;  
using the standard Lax representation 
based on the generators of the Clifford algebra, 
one can easily solve the Kulish--Sklyanin model by applying 
the inverse scattering method or the Darboux transformations. 
However, the obtained exact solutions such as the $N$-soliton 
solution naturally 
involve the generators of the Clifford algebra 
satisfying the anticommutation relations 
and thus are 
not so 
useful 
for further 
analysis. 

In this paper, we 
translated 
the 
standard Lax representation 
for  
the Kulish--Sklyanin model
into the nonstandard one, not involving 
the generators of the Clifford algebra, 
and then 
applied the 
binary Darboux transformation. 
The $N$-fold 
binary Darboux transformation can also be formulated in 
simple explicit 
form, so 
we 
could 
obtain 
a 
classical expression 
for the general $N$-soliton solution 
of the 
Kulish--Sklyanin model (\ref{reducedKS2}),
which is more useful for further investigation. 
By changing the time dependence of the linear eigenfunctions 
and considering a natural reduction, we could also 
obtain a general formula for 
the 
multisoliton (or multi-breather) solutions 
of the vector mKdV equation 
(\ref{vmKdV3}); 
by imposing some additional conditions, 
we obtained  
the 
real
$N$-soliton solution of the vector mKdV equation.

\end{document}